\documentclass[aps]{article}
\usepackage{geometry}                
\usepackage{caption}
\usepackage{subcaption}
\usepackage{graphicx}

\usepackage{amssymb}
\usepackage{epstopdf}
\usepackage{amsmath}

\usepackage{amsfonts}
\usepackage{bm}
\usepackage{enumitem}
\DeclareMathAlphabet{\mathpzc}{OT1}{pzc}{m}{it}

\begin{document}

% Use the \preprint command to place your local institutional report number 
% on the title page in preprint mode.
% Multiple \preprint commands are allowed.
%\preprint{}

\title{Exactly Solvable Dielectrics, Radiation Induced Forces and Causality} %Title of paper

\author{Clifford Chafin\\\ \small{Department of Physics, North Carolina State University, Raleigh, NC 27695} \thanks{cechafin@ncsu.edu}}

\date{\today}
\maketitle

\begin{abstract}
We present an exactly solvable model of a classical dielectric medium that gives an unambiguous local decomposition of field and charge motion and their contribution to the conserved quantities. The result is a set of four branches to the dispersion law that gives full independent freedom in the selection of initial data of the fields and charge motion, in contrast with constitutive laws.  This is done with special care to the forces that exist at surfaces, coatings and the ends of packets.  As a result the utility of a stress-tensor as a function of field strengths and dielectric response for deriving general forces is called into question.    The Abraham-Minkowskii paradox is clarified from this point of view and the export of such notions to realistic media and metamaterials are discussed. One result of this model is a mathematically simpler and more intuitive understanding of causality in media than the Brillouin and Sommerfeld theories. Necessary elastic medium response is estimated and some implications of this picture for quantum effects are included based on conservation laws.  This model can be extended to manifestly maintain these features as general nonlinear and time and space dependent changes in medium response are introduced.  The extent to which this can provide a universal description for all dielectrics is discussed.  A microscopic treatment of negative index materials from this point of view is included as an illustration of the extreme economy and simplicity of these methods.  
\end{abstract}

%\pacs{}% insert suggested PACS numbers in braces on next line

%\maketitle %\maketitle must follow title, authors, abstract and \pacs

% Body of paper goes here. Use proper sectioning commands. 
% References should be done using the \cite, \ref, and \label commands
The Abraham-Minkowskii paradox has led to a century long debate on the proper assignment of momentum to electromagnetic waves in media \cite{Pfeifer}.  Minkowskii argued that the momentum of a photon should be $p=\frac{n\hbar\omega}{c}$ and Abraham as $p=\frac{\hbar\omega}{nc}$.  Some now argue that this is now resolved and both assignments are correct when boundary conditions are correctly applied \cite{Pfeifer}.  The purpose of this article is not so much to dispute that position but to view this problem and general dielectric response from a different and more intuitive point of view.  This will generally support a unique decomposition of momentum into field and material components but also ultimately challenges the universality of such stresses as a function of the dielectric response.  The physical questions we can ask of an electromagnetic wave interacting with a dielectric medium are 1.\ What are the forces at the surfaces?  2.\ What stresses exist due to a steady state beam traversing the medium?  3.\ What are the impulses (and transient losses) associated with a wave packet traversing the medium?  4.\ How many photons are involved and how does this change at boundaries?  5.\ What are causal, damping and nonlinear (and nonlocal) effects on propagation?  To be able to answer such questions, hopefully easily and with physical intuition, is a sign that we truly do understand the system and are in a good position to extend the theory to more challenging configurations that can arise from nonlinearity, confined geometries and quantum effects.

In this article, we will see that the energy and momentum in the photons in a particularly simple medium can be unambiguously specified and separated from that in the charge and core motions.  Internal stresses can be identified with phase shifts/time lags at the charges and induced standing waves in the medium.  The phase velocity turns out to be the rate at which the spatially oscillating part of the energy density advances.   
The group velocity will be the velocity that the total energy of the packet, including the energy of the charges, propagates which will involve conversion of charge kinetic and potential energy at the ends.  In the nondissipative case, the resonant limit gives diverging stress and ratio of charge energy to electromagnetic wave energy.  Elastic medium response can absorb large fractions of the electromagnetic momentum that is typically returned when the field leaves the medium but, for rapidly changing packets, intense fields or large media, the induced acoustic motions can drain energy from the waves in an unrecoverable fashion.  This is just one sort of nonlinearity that media can display and will lead us to a general discussion of nonlinearity and causality in media from a physical point of view rather than the formal approaches used in the Kramers-Kronig relations.  

One might wonder why one should pursue such a reformulations of the established treatments of dielectric response.  One reason is the frustrating and persistent problems that pseudomomentum and pseudostress introduce into physics \cite{McIntyre}.  Even those bent on resolving these conflicts often add new elements to the confusion.  By providing a system that can be exactly parsed locally in terms of conserved quantities one can clarify exactly where the real momentum is, how it is transferred and what the possibilities are for real stresses.  Additionally, there is pedagogical value.  One should not confuse agreement with experiment as a correct derivation.  It is often joked that graduate students can derive whatever you ask them to by some means.  However, none of us are beyond finding overly clever and hopeful derivations based on pleasant seeming abstraction to obtain a known or suspected result.  To be able to attack a problem from multiple directions is a great comfort that our results are valid even when some arguments seem too clever and might belong more in the domain of mathematics than physics.  

Many students express frustration that the treatment of dielectric response suddenly involves the tools of complex analysis, often their first introduction to it, and wonder why an explanation must require such a formal approach.  Physicists and mathematicians may differ on what the more ``fundamental'' sort of treatment means.  Most physics students would consider a microscopic description of the medium to be the most illuminating and natural since it has a clear connection to the fundamental laws of motion.  However, the complexity of such treatments is generally out of reach of their time and ability and have their own approximations that leave room for doubt.  Furthermore, one has the feeling that this is a simple classical problem by nature and it should not require hard tools or wild leaps from the case of single driven oscillators to have an intuitive answer to medium response.  

In the following, we will introduce an exactly solvable model that shows that the energy transfer entirely consists of the electromagnetic flux where the ends of packets act as sources and sinks of this in terms of charge kinetic and elastic bond energy.  This then gives a very nice consistency condition with conservation laws and the allowed dispersion relations that are related to a similar result for surface waves, a much more complicated system.  Delightfully, this approach is not just intuitive but also mathematically rudimentary, brief and leads to some new insight on the ways nonlinearity must essentially enter these systems.  In this model, momentum effects can be exactly separated and classified as: reflection impulses, pure electromagnetic flux, longitudinal plate momentum, radiative stress, forces from the static/velocity fields and bond distortion and conversion impulses due to the ``rings up'' time of the medium to create an equilibrium state with the wave.  During this ring-up, the medium is absorbing energy (hence momentum) from the propagating wave so that the forces on it are not a simple function of the instantaneous fields which means deriving the forces from a stress tensor built from them must fail in general.  We will see that the rest of the internal stress in the medium is due to a hidden standing wave component between the layers due to a phase shift induced by the charges.

The linearized theory of medium response gives a consistent causal theory even though some aspects of induced stress and conservation laws are obscured.  How to handle nonlinearities as extra interactions on a basis built from a linearized theory is not always obvious.  Formal procedures exist in the form of perturbation theory and path integrals  \cite{Hasselmann} but the domain where this is valid and applicable is not always clear.  As an example, the case of water surface waves is quite troublesome.  A broad enough spectral distribution can lead to local fluctuations large enough to induce surface breaking which introduce vorticity and remove energy, momentum and angular momentum from the waves and convert it to heat, nonirrotational flow and angular momentum of the gravitational source.  The end-of-packet contributions of optical signals in media will be seen to necessarily give nonlinear contributions as a function of the medium's elasticity independent of the extent to which the restoring forces are becoming nonlinear.  The internal properties of a medium can easily be induced to change rapidly in real time so that no well defined basis of eigenstates, even including acoustic response, can span the solution.  
To this end, we will present a general local theory that includes damping and nonlinearity for which causality manifestly holds. (Nonlocality due to a sparse phonon spectrum will not be included since this will create a kind of incoherence among photons that is not describable by a classical theory.)  This allows the introduction of any local damping law and medium response restricted only by conservation laws that are clear from the initial form of the equations.  

The organization of the paper is as follows.  In Sec.\ \ref{comment} we comment on the Lorentz-Drude model and the vagueness in it that makes extracting forces from it difficult.  In Sec.\ \ref{exact}, we create an exactly solvable model that gives these results in the nondissipative limit in a fashion that allows an exact decomposition and tracking of energy and momentum as it moves through the medium.  This allows us in Sec.\ \ref{impulseandstress} to explicitly describe the forces that must exist at boundaries, ends of packets and the time delays that must occur in energy transfer from radiative field to medium.  Additionally, we can estimate acoustic losses to the medium, photon number changes in the medium (from purely classical considerations!) and forces on anti-reflective coatings.  In the latter case, we exploit that the momentum flux conservation implies the result must be universal for any medium and show that this can be calculated with extraordinary economy by introducing magnetic monopole currents.  In Sec.\ \ref{damping}, we show how this model can be extended to introduce damping and nonlinear effects which completely respect causality and then show that this model is not completely universal but is, in some sense, minimal in its account of energy and forces in the medium.  An appendix covers the extension of the model to ``left handed'' materials, where group and phase velocities are opposite.  The radiation field is shown to consist entirely of photon fields that move opposite to the phase velocity of the electric field vector that is smoothed over many radiator spacings.  Another appendix discuss causality in diffusive systems as another example of how correct small scale accounting fixes the apparent superluminal problems introduced by continuum mechanical limits.  The final appendix discusses some subtle points on stress tensors and why they may be of limited value in describing the forces on an optical media in the presence of radiation.

%deformable, noninertial

\section{Comments on Classical Constitutive Theory}\label{comment}

Generally, discussions of theory of dielectric response begin very formally and derive the Lorentz-Drude model by introducing a complex dielectric function that gives an out-of-phase damping term.  In real space this corresponds to a spatially and temporally local damping term.  Often there is an appeal to the transfer functions of a localized driven damped oscillators as a strong analogy.  However, the driving and damping are due to fields that are being changed by the motion of the charges and it is easy to get lost in the rather formal definitions of the ``macroscopic'' variables $D$ and $H$.  

If we were to construct a complete basis of the system one might wonder how there can be any damping at all.  The radiational degrees of freedom combined with the electron oscillations and core vibrations are all that exist in the theory.  Quantum statistical mechanics has never adequately reconciled this problem and the Kubo formula is a formal approach to derive results \cite{Chafin-thesis}.  Classical electrodynamics is the coherent limit of quantum electrodynamics.  Losses can take the form of a transition to fields and crystal and collective electronic oscillations that have no classical meaning.  This suggests that the losses that we describe with the imaginary part of the dielectric constant have a purely quantum meaning (in that they relate to incoherent motion with correlations outside of classical descriptions).  %This makes the classical inclusion of damping an essentially ad hoc process but we will find a basis of temporally damping spatially harmonic solutions that are consistent with Lorentz-Drude based on a local sink of energy that is linear in the charge displacement velocity.  %One cannot preclude variations in this based on the temperature of the medium or transverse losses out of the medium that are neglected by the description of a signal as a coherent beam.  If these can be described locally and linearly they must give similar results.  

There is a long history behind the differences between $E,B, D$ and $H$ and which are viewed as fundamental \cite{Purcell}.  Originally, $E, H$ were considered fundamental because of our use of magnets to generate fields.  Now we consider $E,B$ as the fundamental microscopic fields and $D, H$ as some measure of their macroscopic response (although more general mixing of linear responses than this are possible). We will confine ourselves to the electric case.  

In the case of electrostatics, we define the displacement vector $D=\epsilon_{0}E+P=\epsilon E$ where $\epsilon=\epsilon_{0}$, the ``permeability of free space'' for vacuum and larger values for media.  This quantity is chosen for the property that $\nabla\cdot D=\rho_{f}$ so that only the free charges act as sources.  In general, solving for the electric field and polarization of the medium would require an iterative self-consistent approach of finding the polarization including the fields from the surface and other uncanceled fields from internal bound charges.  The use of $D$ allows many highly symmetric problems to be quickly solved by boundary condition constraints and special functions.  We can show that the internal energy density stored in the material is $\frac{1}{2}D\cdot E$.  Beyond this, its meaning is unclear.  It is certainly not the local spatial average of the electric field in a medium.  It might best be thought of as an intermediary step to finding the polarization as $P=(\epsilon-\epsilon_{0})E$ which is a more physically meaningful quantity.

When we seek a response to a time changing field, we generally elevate the dielectric constant to a function of frequency: $P(\omega)=(\epsilon(\omega)-\epsilon_{0})E$.  This implies that 1.\  there has been a relaxation of the medium to a state where $P$ and $E$ obey a constitutive relation (and there is only one such branch for a given $\omega$) and 2.\  harmonic motion exists as solutions and linear combinations of these give general solutions.  We know that electrostatics is not the low frequency limit of electrodynamics.  (Note that E and B fields must both coexist in electromagnetic waves as $\omega\rightarrow0$.) 
Nonlinear effects at the edges of packets appear which are essential to any discussion of the Fourier transformed fields and media response when it comes to momentum conservation.  Linear combinations are limited in their ability to capture this aspect of the physics.  While these nonlinear effects can be locally made arbitrarily small by gentler packet gradients, the contributions are additive so cannot be neglected this way.  This suggests we will ultimately need to work with purely real space fields to answer such questions thus limiting the value of working with the eigenstate basis.  

The extension of the permittivity to complex values is done to consider linear responses that include dissipation.  This could equivalently be done with a real response function that is just $90^{\circ}$ out of phase from the electric field.  This distinction matters because extension to the nonlinear domain is not necessarily able to be done using complex fields where real parts are later taken.  We won't be interested in such strong fields for this paper but when the nonlinearities are very small there are some simple workarounds \cite{Aspnes}.  

The Kramers-Kronig relations assume that the general response function is in this linear domain \cite{Jackson}.  The assumption of causality used in this derivation is not the relativistic one but a local one considering the polarization as response of the driving electric field and that this ``response'' temporally follows the driving.  The motivation of this derivation seems to be the response function of a driven damped oscillator.  Such an oscillator is a spatially localized system where no space-time relativistic causality problems enter i.e.\ there is no evolving ``front'' to observe.  Radiation has this as an intrinsic feature and the response radiates out from each point.  Furthermore, these fields are constantly getting absorbed and reemitted by radiators to which the ``driver''  of the response, medium or field, is ambiguous.  First, we will consider a dissipationless continuum model of an electromagnetic wave in a medium which makes no such distinctions and incorporates the full degrees of freedom available to the system then consider damping effects later.  

\section{An Exactly Solvable Model}\label{exact}

We now seek an exactly solvable model based on an idealized solid.  Realistic solids are composed of many atoms with essentially fixed cores and outer electronic shells that can oscillate.  The actual displacements of these electrons at microwave frequencies and higher is very small for almost all typical radiation strengths.  This is why the linear regime dominates.  Even the nonlinear regime generally exhibits relatively small signals and this is often treated by quantum optical methods where the equations again become linear.  

\begin{figure}
 \includegraphics[width=5in,bb=0 0 1000 1000]{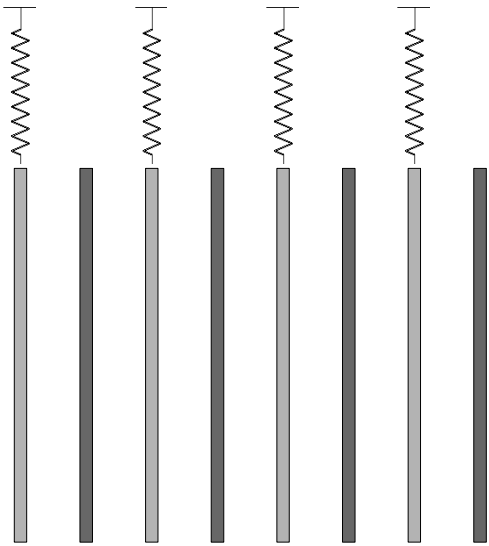}%
 \caption{Alternating positive and negative plates of mass $dm$ and charge $\pm dq$.  The negative plates are attached to springs with constant $d\kappa$.  The positive plates are fixed by constraint.\label{plates}}%
 \end{figure}

The problem that we run into in modeling the field in realistic solids is that there is already a standing electric field from the cores holding the electrons in place at equilibrium.  This has a complicated structure even for a crystal.  On top of this is the radiation field that will be passing through with possible evanescent local contributions.  The Clausius-Mossatti relation gives an expression that relates the local atomic polarizability to the mean dielectric response of the medium.  As the wavelengths in the medium become shorter and approach the interparticle separation, the derivation of it becomes less convincing.  To get around these complications we introduce a model made of vertical charged plates so that we have a 2D translational symmetry in our solutions.  As a first model we choose the arrangement in Fig. \ref{plates}.

\begin{figure}
   \centering
   \includegraphics[width=5in,bb=0 0 1000 1000]{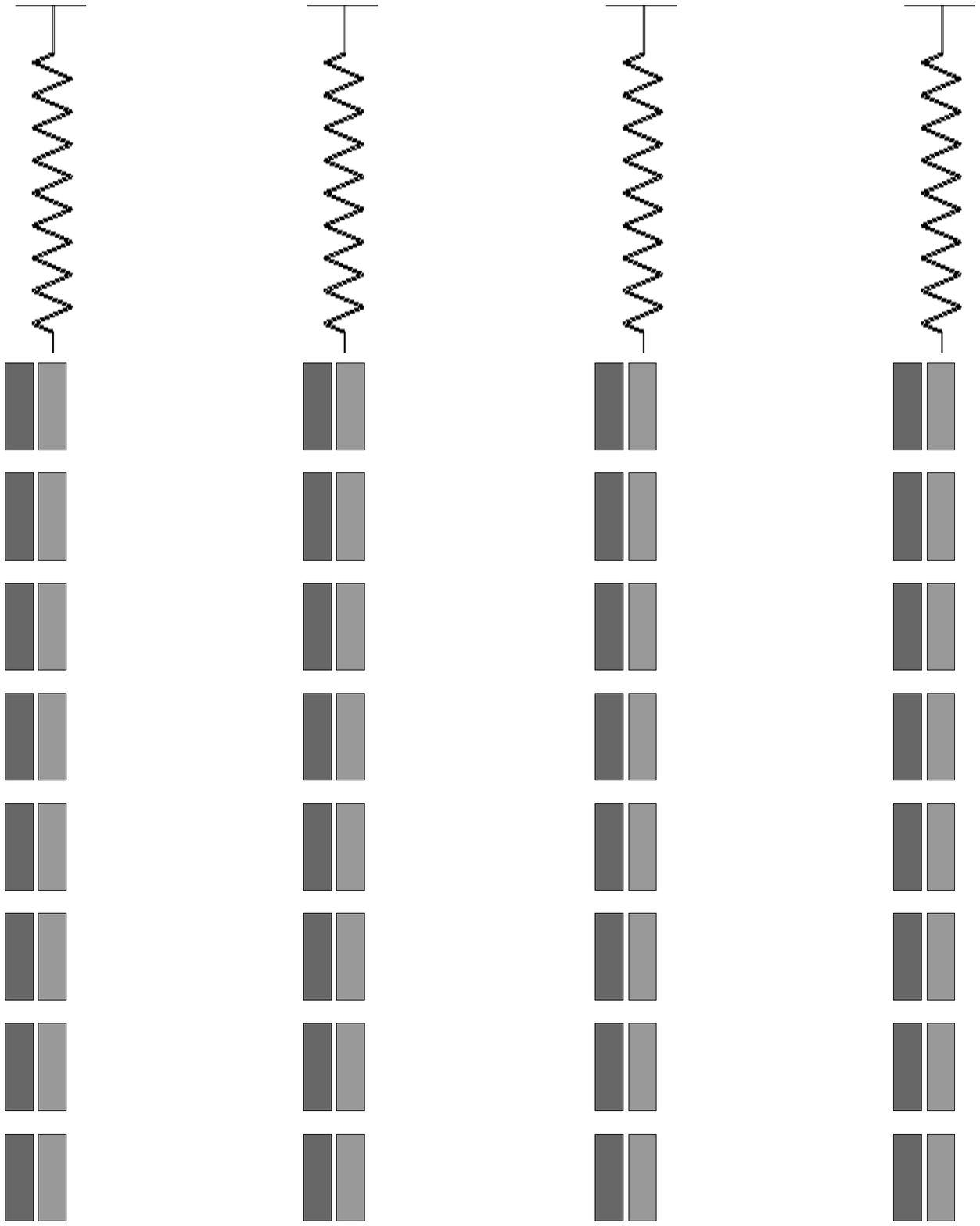} 
   \caption{Close plate pairs.}
   \label{plates1}
\end{figure}

This has a restoring force given by the springs with constant $d\kappa$.  There is some ambiguity as to the state of the static electric field between the plates.  The plate pairs have alternately uniform field and zero field between them.  This inhomogeneity does not seem very physical.  
Closer inspection reveals that displacement of the plates gives strong fringe fields. These will contribute additional forces to restoring the plates.  It is however causally problematic to have spring forces and fields playing a role at distances to the plates longer than $\lambda$.  We are looking to mock up the role of restoring forces in a realistic solid.  These are due to the deformation energy of deformed orbitals and the static electric fields of the cores on the electrons.  For this reason we consider the modified plate arrangement as in Fig. \ref{plates1}.

This confines the fields between the plates so the large gap regions between them contain no field.  We interpret the springs to be acting locally.  The fringe fields that occur from displacements are spread among the many gaps in the material on a scale much less than $\lambda$ of any to-be-considered radiation field.  The restoring forces from displacements are due to the springs and the displacement fields.  These we combine into a net effective force constant $d\kappa$.  Note that we have not taken the dielectric constant $\epsilon$ or index of refraction $n$ as basic here.  The mass and charge densities per area, $(\sigma_{m}, \sigma_{e})$, of the moving charges and the elastic response density per area $K\sigma$ are the primitive microscopic descriptors of the medium.  Note that the ratios of of surface densities per plate and corresponding ratios of volume densities, $(m\rho, q\rho, K\rho)$ with $\rho$ the density of oscillators, will be the same regardless of $d$, the plate separation since they each satisfy relations of the form $\rho=\sigma/d$.   

The advantages of this model are both its symmetry and its ability to let us separate the radiational and nonradiative internal fields in a convenient way.  Because of the symmetries of the system, the values of the fields are unambiguous in between the plates.  The parcel averaged net field in a medium is not obviously of much value and one has to wonder if the fields will have different values in different media with the same macroscopic dielectric properties.  If so, conservation laws may be the only universal way to describe such systems.   The energy is a quadratic function of the fields so using the regional average of a strongly changing field can miss the correct energy by a large amount.  (We often talk about dielectric response as linear and, in the sense that the elastic response of the charges is in the linear regime it is, but it is not meant to imply that the fields are only slightly changed from the vacuum values.  The changes can be quite large and vary rapidly over short distances.)  In our model, we never need to consider these restoring fields even when radiation is passing through the system.  In this case, they are not static but restoring fields yet, assuming the vertical gaps and their separation between each other on the same plate is much smaller than the separation of the plate pairs, they still are clearly distinguishable from the radiation field in these gaps.  In the case of a boosted medium the restoring field picks up magnetic components but the decomposition between these two types of fields is unchanged.  This  facilitates a simple consideration of the form of the final dielectric response functions under relativistic transformations.

\subsection{Equations of Motion}
Averaging methods have the problem that the medium is held up by a balance of electrostatic attraction and wavefunction curvature where the unshielded fields inside the electron shells become extremely large.  The internal field of a disturbed solid has both radiative and velocity fields, only the former of which we tend to think of as radiation.  However, when we try to compute how the energy in the system is stored and the response we do it in terms of the field strengths and it then becomes ambiguous if we should use the net or some nonradiationally subtracted local fields to average and if we should average over all of space or to subtract some interior region of the atoms.  When it comes to the phase of the radiation it will get strongly distorted from a well defined plane wave on the scale of the atoms.  

\begin{figure}[h]
\centering
   \includegraphics[width=3in,trim=10mm 90mm 10mm 80mm,clip]{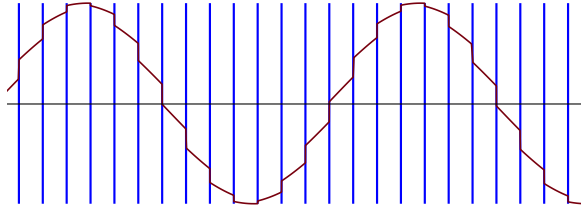} 
   \caption{ Free waves advancing with phase shifts at layers.  The direction of motion of the crests of this smoothed net wave need not be the same as the corresponding crests of the free electromagnetic waves between the layers.  This represents either the magnetic field component or the electric field minus the standing wave component.}
   \label{wavesteps}
\end{figure}

In this model system, the symmetry of the problem allows a precise decomposition $E_{\text{Net}}=E_{\text{restoring}}+E_{\text{rad}}$  and we can give equations of motion for the $E_{\text{rad}}$ henceforth to be called simply $E$.  The phase distortion is replaced by a discrete jump and the singular charge surfaces so that the wavelength and frequency can be well defined between the plates even when the separation $d<<\lambda$ (For example, see fig.\ \ref{wavesteps}).  
Let us generally assume this plate separation is much smaller than the wavelength $\lambda$ of a wave in this vacuum cavity between the plates.  The displacement of the charged plates from equilibrium will be labelled $Y$ so that the polarization density is $P=q\sigma Y$.  The current density is related to $Y$ through $J=\dot{P}=q \sigma \dot{Y}$.  

We assume that a sinusoidal wave is propagating in the x-direction with E-field polarized in the y-direction.  The equations of motion are 
\begin{align}
\label{charges}
dm\cdot\ddot{Y}=de\cdot E-dk\cdot Y
\end{align}
where $dm=m\sigma A$, $dk=K\sigma A$ etc.\ for a plates of area $A$.  

Assuming a sinusoidal solution for the electric field at a given location $x=0$ (so we can ignore any $k x$ terms for now), $E=E_{0}\cos(-\omega t)$ and $Y=Y_{0}\cos(-\omega t)$ we find
\begin{align}
Y(t)=\frac{q E_{0}}{K-\omega^{2}m}\cos(-\omega t)
\end{align}
The current density 
$\mathbf{J}$ is
\begin{align*}
\mathbf{J}&=q\rho\dot{Y}~\hat{\bf{y}}\\
&=\frac{q^{2}\rho\omega E_{0}}{K-\omega^{2}m}\sin(-\omega t)~\hat{\bf{y}}
\end{align*}
We can now apply Maxwell's equations to obtain
\begin{align*}
\nabla^{2}\textbf{E}=\mu_{0}\dot{\bf{J}}+\frac{1}{c^{2}}\partial_{t}^{2}\bf{E}
\end{align*}
and using the full space and time dependent ansatz $\mathbf{E}=E_{0}\cos(k x-\omega t)\hat{\bf{y}}$ we find the dispersion relation (see fig.\ \ref{dispersionrelation}):
\begin{figure}
   \centering
   \includegraphics[width=3in,trim=10mm 100mm 10mm 10mm,clip]{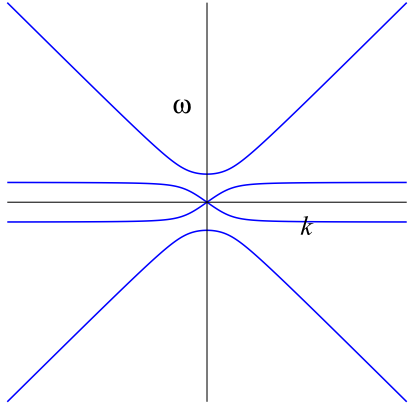} 
   \caption{The dispersion relation. }
   \label{dispersionrelation}
\end{figure}

%\begin{widetext}
\begin{align}
\label{dispersion}
k^{2}=\mu_{0}\frac{\rho q^{2}\omega^{2}}{K-\omega^{2}{m}}+\frac{1}{c^{2}}\omega^{2}=\frac{\omega^{2}}{c^{2}}\left(1+ \frac{\rho}{\epsilon_{0}}\frac{q^{2}}{K-\omega^{2}m} \right)
\end{align}
%\end{widetext}
 gives a phase velocity of
\begin{align*}
v_{\text{ph}}=\frac{\omega}{k}=\frac{c}{\sqrt{1+\mu_{0}\rho {q}^{2}c^{2}\left( {K}-{\omega^{2}{m}}\right)^{-1}}}
\end{align*}
and, by the definition $v_{\text{ph}}=c/n$ and index of refraction
\begin{align*}
n&=\sqrt{1+\frac{\rho {q}^{2}}{\epsilon_{0}({K}-{\omega^{2}{m}})}}
\end{align*}
which is often written in terms of the plasma frequency: $\omega_{p}^{2}=\rho q^{2}/m\epsilon_{0}$ so that
\begin{align*}
n&=\sqrt{1+\frac{\omega_{p}^{2}}{({K/m}-{\omega^{2}})}}
\end{align*}

For completeness we should find the B-field and show consistency of the fields solution.  Note that since E and B are perpendicular to the direction of propagation, $\hat{\mathbf{x}}$, and the plates are constrained to move in this plane, we do not need to worry about $ v\times B$ forces.  These forces are periodic, so cancel \cite{note}, 
 are additionally an order of $c$ smaller than those of $E$ but when we wish to consider coupling to phonon oscillations of the medium and damping we should include them.  (In the case of compact packets these magnetic forces will impart net end-of-packet impulses that distinguish between the Abraham, total packet including electronic and core, and Minkowskii, local internal electromagnetic, definitions of momentum.)

First we solve for the B-field:
\begin{align*}
\nabla^{2}\mathbf{B}+\mu_{0}\nabla\times\mathbf{J}=\frac{1}{c^{2}}\partial_{t}^{2}\mathbf{B}
\end{align*}
Assuming $\mathbf{B}=B_{0}\cos(kx-\omega t)\hat{\mathbf{z}}$ we find
\begin{align*}
k^{2}B_{0}=\mu_{0}\frac{\rho {q}^{2}\omega k}{K-\omega^{2}{m}}E_{0}+\frac{1}{c^{2}}\omega^{2}B_{0}.
\end{align*}
With $E_{0}=\frac{\omega}{k}B_{0}$ we get consistency with the equation of motion for $\bf E$ and the original Maxwell's equations.  Collecting our solutions we obtain: 
\begin{align}
\mathbf{E}&=E_{0}\cos(k x-\omega t)\hat{\bf{y}}\label{solution}\\
\label{B}
\mathbf{B}&=\frac{n}{c}E_{0}\cos(k x-\omega t)\hat{\bf{z}}\\
\label{Y}
\mathbf{Y}&=\frac{ qE_{0}}{K-\omega^{2} {m}}\cos(k x-\omega t)\hat{\bf{y}}\\
\mathbf{J}&=\frac{\rho {q}^{2}\omega E_{0}}{K-\omega^{2}{m}}\sin(k x-\omega t)~\hat{\bf{y}}
\label{solution1}
\end{align}
with the dispersion relation $k=n\omega/c$ where $n$ is given above.  

Eqn.\ \ref{dispersion} gives a set of four branches of $\omega$ for each $k$.  Two are the usual radiationally dominated modes that have $n(k)\rightarrow1$ as $k\rightarrow\infty$.  There are two other modes that have $\omega\rightarrow0$ as $k\rightarrow0$.  These we might be inclined to call ``acoustic'' modes in analogy with similar results in condensed matter theory.  In contrast with the first two modes these are ``elastically dominated'' (to be made more clear later).  If we let the charge density vanish, these modes degenerate to a set of independent oscillators with frequency $\omega_{0}=\sqrt{K/{m}}$.  The phase and group velocity of these modes are all less than $c$ for the acoustic branch and no singularities occur.  However we notice the singular denominators in $Y$ and $J$ corresponding to $\omega=\omega_{0}$ the resonant cutoff of the acoustic modes where $k$ diverges.  This means a packet would usually contain negligible amounts of such modes.  
These give a complete set of variables that allows us to specify any state of the field and medium.  Presumably we could expand any packet with vanishing fields, displacements and currents outside a compact set and observe the advance rate of the edge.

Before we move on to more subtle considerations, let us address the question of group velocity and the energy and longitudinal momentum transfer in the medium.  Our basis did not allow any longitudinal momentum as the plates were constrained in this direction.  This momentum is shared with the elastic modes of the medium which are often very small. However the duration of a packet can be quite long and variations in its intensity can be on periods where acoustic oscillations can be excited. Our basis requires we include these as an ad hoc modification that connects to these acoustic modes to conserve this momentum.  It is this consideration that will lead to an understanding of the physical meaning of the Abraham and Minkowskii momenta for this model and will show a very nice built in self-consistency to the theory and a transparency of how momentum and energy are shared between medium and fields and a causal correction for nonlinear and varying damping effects.  

\subsection{Energy, Momentum and the Group Velocity}

From this model we can now compute the local energy density and momentum directly for the progressive wave solutions.  The energy has three sources
\begin{enumerate}[nolistsep]
\item KE of plates
\item PE of springs
\item Energy of fields
\end{enumerate}

This gives the total energy density in Eqns.\ \ref{Edecomp}-\ref{Edecomp1}
%\begin{widetext}
\begin{align}
\label{Edecomp}
\mathcal{E}&=\frac{1}{2}m\rho\dot{y}^{2}+\frac{1}{2}K\rho y^{2}+\frac{1}{2}\left( \epsilon_{0}E^{2}+\frac{B^{2}}{\mu_{0}} \right)\\
&=\mathcal{E}_{q}+\mathcal{E}_{\text{em}}\\
\label{oscillations}
&=\frac{1}{2}{E_{0}^{2}}\left(\frac{\rho q^{2} K}{(K-m\omega^{2})^{2}}\cos^{2}(kx-\omega t)  +\frac{\rho q^{2}m\omega^{2}}{(K-m\omega^{2})^{2}} \sin^{2}(kx-\omega t)  \right)\\
&\hspace{2cm}+E_{0}^{2}\left(\epsilon_{0}+\frac{\rho q^{2}}{2(K-m\omega^{2})}   \right)\cos^{2}(kx-\omega t) \\
&=E_{0}^{2}\left(\epsilon_{0}+\frac{\rho q^{2}}{(K-m\omega^{2})}\right)\cos^{2}(kx-\omega t)+\frac{1}{2}E_{0}^{2}\frac{\rho q^{2}m \omega^{2}}{(K-m\omega^{2})^{2}}\\
\label{oscillations1}
&=E_{0}^{2}\epsilon_{0}n^{2}\cos^{2}(kx-\omega t)+\frac{1}{2}E_{0}^{2}\frac{m \omega^{2}}{K}\epsilon_{0}(\eta^{2}-1)\\
\label{Edecomp1}
&=\mathcal{E}_{\text{oscill}}+\mathcal{E}_{\text{static}}
 \end{align}
%\end{widetext}
where 
\begin{align}
\eta=\sqrt{\left(1+\frac{K\rho q^{2}}{(K-\omega^{2}{m})^{2}\epsilon_{0}}\right)}
\end{align}
It is related to the index of refraction by
\begin{align}
\frac{ K \epsilon_{0}}{\rho q^{2}}(n^{2}-1)^{2}=\eta^{2}-1
\end{align}
The charge and electromagnetic energy is decomposed here to show that one can view it as static and moving oscillatory components.  This will be useful in later consideration of the phase velocity.  We notice that the oscillatory components of the charge energy is exactly equal, and in phase with, the energy of the electromagnetic waves while the potential energy contribution is out of phase with them leading to a net constant energy density component.  

 The time averaged energy is 
 \begin{align*}
 <\mathcal{E}>=\frac{1}{2}\epsilon_{0}\eta^{2}E_{0}^{2}
\end{align*}
Using the polarization vector defined by $P=q \rho Y=\chi E$ we have $\chi=\frac{\rho q^{2}}{(K-\omega^{2}{m})}$ so that 
 \begin{align*}
 <\mathcal{E}>&=\frac{1}{2}\left(\epsilon_{0}+\frac{K\rho {q}^{2}}{(K-\omega^{2}{m})^{2}}\right)E_{0}^{2}\\
 &=\frac{1}{2}\epsilon_{0}\left(1+\frac{K}{\epsilon_{0}(K-\omega^{2}{m})}\chi\right)E_{0}^{2}\\
 &=\frac{1}{2}\epsilon_{0}\left(1+ \tilde{\chi}\right)E_{0}^{2}\\
 &=\frac{1}{2}\tilde{\epsilon}E_{0}^{2}\\
 &=\frac{1}{2}<\mathbf{E}\cdot \mathbf{D}>
\end{align*}
where $D=\tilde{\epsilon}E$ and $\tilde{\chi}=\frac{K}{\epsilon_{0}(K-\omega^{2}{m})}\chi$.  $\mathbf{D}$ and $\tilde{\chi}$ are contrived here to get the usual relations between energy and displacement.

The Poynting vector is given by
\begin{align}
\mathbf{S}:=&\frac{1}{\mu_{0}}\mathbf{E}\times\mathbf{B}\\
=&  \frac{n}{c\mu_{0}}E_{0}^{2}\cos^{2}(kx-\omega t)\hat{\mathbf{x}}\\
=&  n \epsilon_{0}c E_{0}^{2}\cos^{2}(kx-\omega t)\hat{\mathbf{x}}\\
\end{align}
where $n=\sqrt{(1+\chi)}$. 
The transverse work and longitudinal force on the plates are
\begin{align}
W&=\mathbf{J}\cdot\mathbf{E}=\frac{\rho q^{2}\omega}{K-\omega^{2}m} E_{0}^{2}\sin(kx-\omega t)\cos(kx-\omega t)\\
F&=\mathbf{J}\times \mathbf{B}=\frac{n\rho q^{2}\omega}{K\omega^{2}m} E_{0}^{2}\sin(kx-\omega t)\cos(kx-\omega t)\hat{x}
\end{align}

If we now consider time averages of these quantities we obtain:
\begin{align}
\label{energy}
<\mathcal{E}>&=\frac{1}{2}E_{0}^{2}\epsilon_{0}\eta^{2}\\
\label{S}
<S>&=\frac{1}{2}\epsilon_{0}n c E_{0}^{2}\\
<\mathbf{J}\cdot\mathbf{E}>&=0\\
<\mathbf{J}\times \mathbf{B}>&=0
\end{align}

From the dispersion relation we can calculate the group velocity
\begin{align}
v_{g}=\frac{\partial\omega}{\partial k}=\frac{\sqrt{1+\frac{\rho q^{2}}{(K-\omega^{2}m)\epsilon_{0}}}}{1+\frac{K\rho q^{2}}{(K-\omega^{2}{m})^{2}\epsilon_{0}}}c=\frac{n}{\eta^{2}}c
\end{align}
which is always less than c.  

One can show \cite{Balazs} that center of mass and momentum conservation dictate that a packet must move at the group velocity through a uniform medium and carry momentum density $\mathpzc{p}=\mathcal{E}v_{g}/c^{2}$.  This is a universal relation for any disturbance that travels as a bound compact unit and leaves the medium undisturbed after it passes.  This does not preclude stresses at the ends of the packet and complicated interactions with surfaces and momentum transfer at boundaries and through antireflective films as such a packet enters the medium.  

We can verify this theory in this case by noting that 
$\mathpzc{p}=S/c^{2}$ is the momentum density of the electromagnetic field.  Consider that the plates carry no net momentum.  It is not just that they are laterally constrained.   The forces in the x-direction average to zero over a cycle.  This means that $\mathpzc{p}$ is all the momentum density there is to the packet.
$\mathpzc{p}=S/c^{2}$ is the momentum density of the electromagnetic field and since there is no contribution from the electrons due the lateral constraints on the medium and the fact that these forces average to zero for a plane wave, this is the only momentum in the problem.  
This lets us immediately verify 
\begin{align}
\label{fundamental}
<\mathpzc{p}>=\frac{<\mathcal{E}>}{c^{2}}v_{g}.
\end{align}
Next we will consider details of the microscopic solution and how stress builds up in the medium and is shared between the charges and the fields.

\section{Impulse and Stress}\label{impulseandstress}

The Abraham-Minkowskii controversy revolves around the momentum of an electromagnetic wave in media.  The use of momentum flux is common in treatments of elasticity and hydrodynamics but unlike mass flux (i.e.\ momentum density), momentum is not a locally advected quantity (See App.\ \ref{StressMomentum}).  The pressure plays the role of a source and sink and incompressibility introduces the ability to transport momentum over large distances apparently acausally.  From this springs and endless list of sins and accidental successes from the use of pseudomomentum.  For completeness, and to avoid such pitfalls, we seek to understand all the local forces and momentum densities in a system.  These fall into several types: reflection impulses, pure electromagnetic momentum, longitudinal plate momentum (that has been excluded from the degrees of freedom in our basis), radiative stress from ``hidden'' standing waves, forces from the static/velocity fields and bond distortion that we could term electrostriction, and, most subtle of all, conversion impulses due to momentum absorption of the material while it ``rings up'' to an equilibrium state with the wave.  

As mentioned before, a real understanding of the system and a universal theory of dielectrics will involve our ability to keep track of the conserved quantities as they propagate through the medium and exchange forces with it.  We will see below that the stress tensor of the combination of medium and electromagnetic field together is not that illuminating but expressing them separately and including the local momentum transfers between them gives the observable impulse induced changes in the medium.  The use of packets is essential in this process.  Infinite wave trains can provide exactly solvable solutions but, from the standpoint of conservation laws they can harbor hidden inconsistencies.  Two examples are the case of electromagnetic momentum of charged infinite plates in a magnetic field and that of the angular momentum of surface waves.  In the first case, the net momentum is cancelled by fringe fields at infinity, which we might suspect since no composed collection of charges with no initial momentum can acquire any.  In the second case, right moving progressive waves have a ccw angular momentum about the surface, despite cw particle motion but the value of this depends on where boundaries are periodically placed.  Hence the angular momentum density is sensitive to boundary conditions at infinity \cite{Chafin-waves}.  

\begin{figure}
        \centering
        \begin{subfigure}[b]{0.4\textwidth}
                \includegraphics[width=\textwidth]{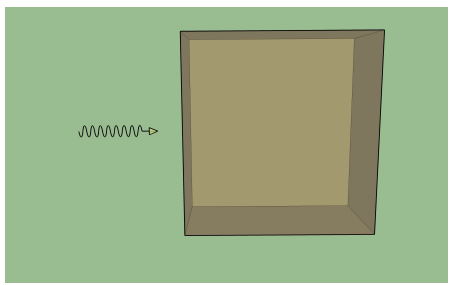}
                \caption{An Incident Packet}
                \label{fig:incident}
        \end{subfigure}%
        ~ %add desired spacing between images, e. g. ~, \quad, \qquad, \hfill etc.
          %(or a blank line to force the subfigure onto a new line)
        \begin{subfigure}[b]{0.4\textwidth}
                \includegraphics[width=\textwidth]{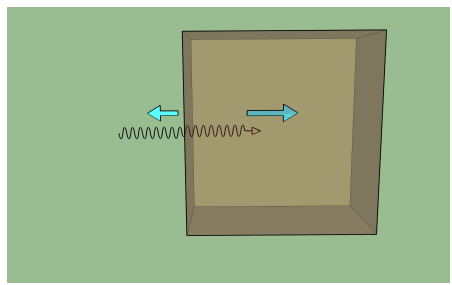}
                \caption{An Entering Packet}
                \label{fig:halfin}
        \end{subfigure}
        \centering
        %        ~ %add desired spacing between images, e. g. ~, \quad, \qquad, \hfill etc.
          %(or a blank line to force the subfigure onto a new line)
        \begin{subfigure}[b]{0.4\textwidth}
                \includegraphics[width=\textwidth]{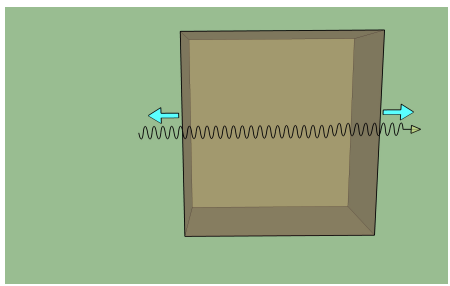}
                \caption{A Traversing Packet}
                \label{fig:traversing}
        \end{subfigure}
        \begin{subfigure}[b]{0.4\textwidth}
                \includegraphics[width=\textwidth]{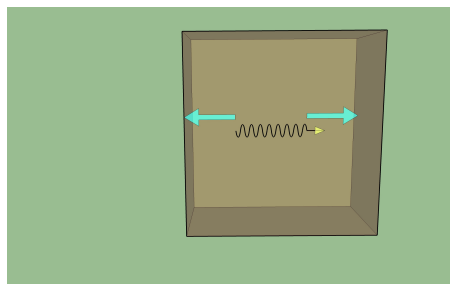}
                \caption{An Entirely Enclosed Packet }
                \label{fig:enclosed}
        \end{subfigure}
        \caption{EM Packets with end-of-packet forces and surface forces indicated by location and length of arrows.}\label{fig:animals}
\end{figure}

In the case of an electromagnetic packet entering a block, fig.\ \ref{fig:incident}, we will see that, even given perfect AR coatings on the surfaces, the medium must acquire part of the momentum of the packet.  There is additionally a stress that exists in the solid at the ends of a packet entirely within the block, fig.\ \ref{fig:enclosed}.  This is, however, different from the stresses that exist at the end of an infinite wavetrain, fig.\ \ref{fig:traversing}, or long packet that traverses the block.  We will compare this with the case of a partly in and partly out packet, fig.\ \ref{fig:halfin}, to demonstrate how the outwards forces on the surface can be present while the medium picks up a net impulse.  We will also investigate the effects of finite time duration of such impulses and how energy can get drained from the packet into acoustic impulses that traverse the block.  

The case of electromagnetic waves in media is complicated by the fact that there are two very distinct components: fields and charges.  In acoustics and hydrodynamics, this is not the case.  Each has other complications that make them more complicated but in this one respect they are simpler.  Progressive surface gravity waves have a number of serious complications not the least of which that they carry mass hence have a nonzero momentum density.  This is generally considered to be a higher order effect and, when it comes to (the lowest nonrelativistic component of) energy transport, it can be ignored \cite{Phillips, Lighthill}.  They will however shed some light on how to consider energy transport and the origins of group and phase velocity from the point of view of conservation laws.  

\subsection{Comparison with Surface Waves}
The small amplitude solution of a deep water surface wave is given by the Airy wave with surface profile $\nu=a\sin(kx-\omega t)$ and velocity potential function $\phi=\frac{\omega}{k}a e^{kz}\sin(kx-\omega t)$ with the following dispersion relation: $\omega =\sqrt{gk}$.   From this we derive the phase and group velocities: $v_{\text{ph}}=\sqrt{g/k}$, $v_{g}=\frac{1}{2}v_{\text{ph}}$.  
The energy density per unit area is $\frac{1}{2}\rho g a^{2}$ where the energy is equally divided among the kinetic and potential energy.  The kinetic energy comes from the the small circular motion that penetrates down to a depth of $\sim \lambda$.  There is a small higher order drift that we ignore here.  Averaged over depth this is a uniform spatial distribution.  The potential energy, however oscillates with position as $\mathcal{U}=\frac{1}{2}\rho g a^{2}\sin^{2}(kx -\omega t)$.  This gives a total energy density of $\mathcal{E}=\frac{1}{2}\rho g a^{2}\{\frac{1}{2}+\sin^{2}(kx -\omega t)\}$.  

The energy density is thus of the form of a constant $\mathcal{E}_{\text{static}}=\frac{1}{4}\rho g a^{2}$ plus a positive advancing oscillatory function of height $\mathcal{E}_{\text{oscill}}=\frac{1}{2}\rho g a^{2}$.  It seems that the potential energy is transporting at velocity $v_{\text{ph}}$ and the kinetic energy is fixed.  The usual interpretation, due to Rayleigh, in terms of packets indicates we should have a flux of $\mathcal{E}_{\text{net}}\cdot v_{g}$ which, interestingly equals $\mathcal{E}_{\text{oscill}}\cdot v_{\text{ph}}$.  When the energy reaches the end of a packet, it produces new elevated regions which do not yet have corresponding kinetic motion to propagate them.  These crests then drive the flows and an analogous effect happens at the back end of the packet.  This gives us a picture of a packet as one where potential energy is transported at $v_{\text{ph}}$ and the ends act as sources and sinks that convert half of this back into kinetic energy.  The kinetic energy in the middle is essentially static (except for a small Stokes drift).  Packet spreading must arise at the ends due to the fact that the pressure created by the surface distribution will extend down and forwards $\sim\lambda$ from the packet's end so always leads to some stretching out of surface elevation of the packet.  

Let us now apply this mode of thinking to our packets of electromagnetic energy in media.  Unfortunately, it is not exactly true that the kinetic and potential energy of the charges is static 
from eqn.\ \ref{oscillations}.  However, we do have a combination of charge and electromagnetic energy that is constant in eqn.\ \ref{oscillations1}.  Here we can decompose the net energy as $\mathcal{E}_{\text{net}}=\mathcal{E}_{\text{oscill}}+\mathcal{E}_{\text{static}}$.  Using $<\mathcal{E}_{\text{oscill}}>v_{\text{ph}}=<S>$ we see that a similar interpretation applies here.  It seems that there is a static component that is not moved by the traveling oscillatory crests.  Unpublished investigation into other systems by the author suggests this is a very general feature when no net mass flux is present.  Since $v_{\text{ph}}$ can be greater than $c$ this is clearly an imperfect interpretation.  Closer inspection shows that there is a back and forth sloshing of the energy that contributes to the oscillating force on the charges.  This oscillating local backwards moving flux of electromagnetic energy explains how the crests of energy above $\mathcal{E}_{\text{static}}$ can advance at greater than $c$ while the total local energy flux never exceeds it.  At this point one might wonder why we cannot simply compute the electromagnetic energy between the plates $\mathcal{E}_{\text{em}}$ and use that it advances at the vacuum group velocity, $c$, so that $S=\mathcal{E}_{\text{em}}c\sim n^{2}$ rather than $n$.  The plates themselves can only transfer energy via the em fields so this should be all of the energy flux.  As we will see in Sec. \ref{Stress} below, the phase shifts at the plates necessitate a microscopic backwards flux of energy for a macroscopically simple progressive wave.  This generates both a stress on the walls of the medium and a reduced effective electromagnetic energy that advances at $c$.

First let us consider the impulses due to the AR coating, specifically, what is the momentum (not internal stress) of these waves in media and what momentum transfer has it exerted on the medium.  (Internal stress is not ignorable here and must be considered in any experimental result.  In this model some of this force is hidden at the far off edges of the block in the fringe fields of our plates. Realistic media will generate this from the local velocity fields and changes in the orbitals due to driving.  Below we just consider the fraction that arises from global conservation laws.)  We will do this two ways.  First by an explicit calculation then by an analysis of the microscopic fields of a wavetrain traversing the medium.

\subsection{Force on an AR layer}\label{AR}
It is tempting to simply use the Lorentz force to compute the forces on the surface using the continuum or constitutive model and be done with it.  Unfortunately, this will not always be sufficient for reasons already suggested.  However, we can directly calculate this force of a steady beam at an AR coating.  
The details of the AR coating is not important, since the energy flux is constant throughout the system, the impulses at it will not be system dependent.  This is in contrast with the stress on it which may be medium dependent due to other sources of electrostriction.  The difficulties with deriving an exactly solvable calculation \cite{Mansuripur} lies in the asymmetry between the two dynamic Maxwell's equations.  The current term $J_{q}$ exists in the $\partial_{t}E$ equation.  We can correct this asymmetry by imposing a magnetic monopole current $J_{m}$.  As long as there is no net work done by these current the time averaged impulse will be unchanged.  The corresponding Maxwell equations are \cite{Moulin}  

\begin{align}
\partial_{t}E&=c^{2}\nabla \times B-\epsilon_{0}^{-1}J_{e}\\
\partial_{t}B&=-\nabla \times E-\epsilon_{0}^{-1}J_{m}
\end{align}
We now match the two solutions with the same phase but discontinuously at $x=0$.  
\begin{align}
E&=E_{0}\cos(\omega t)\hat{y} &   B&=\frac{1}{c}E_{0}\cos(\omega t)\hat{z} & x<0\\
E&=\frac{E_{0}}{\sqrt{n}}\cos(\omega t)\hat{y} &   B&=\frac{\sqrt{n}}{c}E_{0}\cos(\omega t)\hat{z} & x>0
\end{align}
The discontinuities give the implied singular current densities.
\begin{align}
J_{e}&=-\frac{\sqrt{n}-1}{c\mu_{0}}E_{0}\cos(\omega t)\delta(x)(-\hat{y})\\
J_{m}&=\frac{(\sqrt{n}^{-1}-1)}{\mu_{0}}E_{0}\cos(\omega t)\delta(x)(\hat{z})\\
\end{align}
The power can be calculated by dotting the midpoint field at $x=0$ with the singular currents $J_{e}\cdot E_{mid}+J_{m}\cdot B_{mid}=0$.  These vanish since they give opposite contributions.  
The averaged (net outwards) pressure at the surface is given by 
\begin{align}
P_{S}=S_{xx}^{AR}=<J_{e}\times B-J_{m}\times E>=\frac{1}{4}\epsilon_{0}E_{0}^{2}\frac{(n-1)^{2}}{n}
\end{align}
This is consistent with the results of Mansuripur \cite{Mansuripur} derived with considerably more effort.

\subsection{Microscopic Electromagnetic Stress}
\label{Stress}

We next investigate the stress in the medium by investigating the microscopic decomposition of the fields between the plates and show it is consistent with the above result.  Considering the case of fig.\ \ref{fig:traversing}, the medium has completed any relaxation and momentum absorption from the advancing action of the packet edge.

The fields in the gaps are made of plane waves with dispersion $\omega=ck$ where these are presumably of a nearly monochromatic form with the same $\omega$ as the macroscopic frequency.  We can decompose the fields here into right and left moving components by first noting that, for analytic waveforms  with $|E|=c|B|$, we have a traveling wave and removing this component gives a wave moving in the opposite direction.  We start by assuming we have primarily a progressive wave in the x-direction and a standard right handed coordinate system.  In the following $E$ and $B$ are components of these vectors in the induced $(y,z)$ axes.  This gives a decomposition in to right and left moving component waves.  
\begin{align}
E_{\rightarrow}&=\frac{1}{2}E+\frac{1}{2}cB\\
E_{\leftarrow}&=\frac{1}{2}E-\frac{1}{2}cB\\
B_{\rightarrow}&=\frac{1}{2}B+\frac{1}{2c}E\\
B_{\leftarrow}&=\frac{1}{2}B-\frac{1}{2c}E
\end{align}
This decomposition eliminates the cross terms in the stress so that we can write $S=S_{\rightarrow}+S_{\leftarrow}$ in terms of the respective crossed right and left moving fields.  The corresponding momentum densities are:
\begin{align}
<p_{\rightarrow}>=&\frac{S_{\rightarrow}}{c}=\frac{1}{2c^{3}\mu_{0}}\frac{(1+n)^{2}}{4}E'^{2}\\
<p_{\leftarrow}>=&\frac{S_{\leftarrow}}{c}=-\frac{1}{2c^{3}\mu_{0}}\frac{(1-n)^{2}}{4}E'^{2}
\end{align}
Where $E$ has been relabeled as $E'$, the field strength between the plates, a distinction we will need shortly.  

As a consistency check we see that $<\mathpzc{p}>=<\mathpzc{p}_{\rightarrow}+\mathpzc{p}_{\leftarrow}>=S/c^{2}=\frac{1}{2}\epsilon_{0}\frac{n}{c}E'^{2}=\frac{1}{2}\epsilon_{0}\frac{1}{c}E_{0}^{2}$.  This is the net momentum density.  The remainder of the momentum is traveling right and left and canceling but still generating a stress and force on the boundaries of the material.  We can think of this microscopically as the momentum as having a free standing wave component with a free propagating wave between each plate. (An illustration is provided in fig.\ \ref{wavesteps}). The residual flux gets a magnitude of $|c\mathpzc{p}_{\leftarrow}|$ in each direction so an obstruction reflects with twice this magnitude.  (Absorbtion instead of reflection gives half the following result).  The stress is therefore:
\begin{align}
\label{Sxx}
S_{xx}&=<2c\mathpzc{p}_{\leftarrow}>\\
&=\frac{1}{2c^{2}\mu_{0}}\frac{(1-n)^{2}}{2}E'^{2}=\frac{\epsilon_{0}}{4}\frac{(1-n)^{2}}{n}E_{0}^{2}
\end{align}
Note that $E_{0}$ here is the field in the vacuum before the packet entered the medium and $E'$ is the field strength inside the medium.

One could alternately view the role of the plates as inducing a phase shift in the local $E$ and $B$ fields to the extent that they are well approximated as sections of plane waves of wavevector $k=c/\omega$.  Defining the crest maxima as $\theta_{E}$ and $\theta_{B}$ we can define the phase shift $\Delta \theta=\theta_{E}-\theta_{B}$ mod$(2\pi)$.  The shifts at each plate gives an infinitesimal shift in the energy and momentum of the waves.  This gives an alternate perspective to the above point of view in terms of right and left traveling waves.  We can then do a calculation of forces bases on the ``phase shift density.''  (Realistic media have no such singular sheets of charge and this phase shift must be replaced by a stretching of phase near the charges.)  

\subsection{The ``Ring-Up'' Impulse}
\label{Impulse}
Now let us do a careful calculation of the motion of a packet moving into a dielectric slab with perfect AR coatings in the spirit of Balazs \cite{Balazs} where the packet begins as in fig.\ \ref{fig:incident} and arrives on the interior as in fig.\ \ref{fig:enclosed}.  Let the packet have length $L$ (less than the length of the slab), area $A$, and roughly monotonic frequency $\omega$.  Its energy is $\frac{\epsilon_{0}}{2}(A L) E_{0}^{2}$.  The packet moving into the medium advances at $v_{g}$ and is contracted by a factor of $v_{g}/c=n/\eta^{2}$ so, by the above relations has energy $\frac{\epsilon_{0}}{2}(A L \frac{v_{g}}{c}) \eta^{2} E'^{2}$, where $E'$ is the maximum field intensity in the medium vs.\ $E_{0}$ as the field intensity maximum in the vacuum.  
 The yields the relation between the field in vacuum and the medium: $E'=\frac{1}{\sqrt{n}}E_{0}$. Computing the net Poynting vectors of the packet inside and outside we find they are identical: $S=S'=\frac{1}{2\mu_{0}c}E_{0}^{2}$ therefore the momentum densities are also equal $\mathpzc{p}'=\mathpzc{p}$.    
Since the packet gets contracted by $n/\eta^{2}$ we see that there is a deficit of momentum $\Delta p=\frac{1}{2\mu_{0}c^{3}}E_{0}^{2}AL(1-\frac{n}{\eta^{2}})$.  
This force is inwards and is returned later when the packet leaves the medium.  The resulting force depends on the the duration of the packet $\Delta \tau=L/c$.  The induced pressure is 
\begin{align}
\label{impulse}
P_{B}=\frac{\Delta p}{\Delta \tau}=\frac{1}{2\mu_{0}c^{2}}E_{0}^{2}\left(1-\frac{n}{\eta^{2}}\right)
\end{align}

Now consider the evolution of the packet midway entering the slab as in fig\ \ref{fig:halfin}.
There must be a ``ring-up'' time to build up the standing wave and impart energy to the oscillating charges during which time the associated electromagnetic momentum is absorbed by the medium.  We can estimate the lag experienced by the front of the wave.  The energy density of the wave is given by eqn. \ref{energy}
\begin{align}
<\mathcal{E}>=\frac{1}{2}E'^{2}\left(\epsilon_{0}+\frac{K\rho q^{2}}{(K-\omega^{2}{m})^{2}}     \right)
\end{align}
Assuming no reflection from the edge of the advancing front, we find a front velocity that advances at 
\begin{align}
c'=\frac{\mathcal{E}_{\text{vacuum}}}{\mathcal{E}_{\text{medium}}}c=\frac{n}{\eta^{2}}c=v_{g}
\end{align}
which is always $\le c$.

\subsection{Net Stress and Forces}
We can now summarize what we know in terms of forces on the surfaces and packet ends in terms of $P_{S}$ and $P_{B}$.  As the packet enters the medium as in fig.\ \ref{fig:halfin}, there is a backwards pressure on the wall $-P_{S}$ and a forwards pressure on medium at the advancing front $+P_{f}$ such that the net force is $(P_{f}-P_{S})A=P_{B}A$ so that $P_{f}=P_{B}+P_{S}$.  The electromagnetic standing wave is sitting between the back wall and the front edge of the packet so that we can identify $P_{ring-up}=P_{B}$ as the pressure from the ring-up of the medium.  

Once the packet is entirely enclosed as in fig.\ \ref{fig:enclosed} in the medium we have equal forces on the medium on each side due to the stress and ring-up and ring-down respectively: $P=P_{B}+P_{S}$.  In the case of the long traversing wave in fig.\ \ref{fig:traversing} we have only the $P_{S}$ forces at the surface.  This gives us a picture of the equilibration of a long beam as transferring a net impulse to the medium then introducing a net stress across the walls.  For reflections at an immersed reflector, as in the Ashkin-Dziedzic experiment \cite{Ashkin}, the net impulse on it is from the momentum flux of the external beam with no $P_{B}$ forces at all.  This explains why the Minkowskii definition of the momentum works for this experiment.  

\subsection{Acoustic Losses}
We have already discussed how damping in media must be an essentially quantum event since, otherwise, the associated elastic damping modes would simply enrich our band structure and never remove energy from the beam.  However, for the case of modulated beams one can have losses into acoustic modes.  In principle such modes simply give new branches of an otherwise lossless dispersion relation.  When the relative occupancy of these acoustic modes are small one can view them as a kind of sink and this is the approach we take here.  In practice, acoustic modes will be damped to quantum incoherent motions of the media and never recontribute to the beam.
  
As the front of a half infinite beam advances across the medium it imparts a force at this rapidly advancing layer.  The front surface of the block experiences a nearly uniform force $P_{S}A$.  The medium, of length $L$, will equilibrate to this on some time scale $\tau\gtrsim L/v_{s}$ where $v_{s}$ is the speed of sound in the medium and $Y$ is its Young's modulus.  Let us now modulate our beam into a set of pulses with duration $T_{d}$  and spacing $T_{s}$.  The impulses create acoustic stress waves of length $l\sim v_{s}T_{d}$ and amplitude $\xi=\mathcal{A}\sim P_{S}l/Y$.  These move through the medium at $v_{s}$ so that if $T_{s}\gg T_{d}$ these packets are well separated.  The mean averaged energy flux removed by acoustic means is then
\begin{align}
\mathcal{P}_{a}
\sim\frac{ P_{S}^{2}}{v_{s}\rho}\frac{T_{d}}{T_{s}}
\end{align}

This gives an acoustic loss in the beam energy that is not present in the case of a uniform beam traversing the medium and reduces the power flux from $\mathcal{P}_{\text{em}}=\frac{1}{2}\epsilon_{0}cE_{0}^{2}$ to $\mathcal{P}_{\text{em}}-\mathcal{P}_{a}$.  One may wonder if there are acoustic losses in the medium during uniform propagation of a beam.  At zero temperature there will be beam solutions that incorporate acoustic response into them but at finite temperature, thermal fluctuations will remove a fraction of energy \cite{LL-fluid}.  

\subsection{Quantum Considerations}
The classical fields correspond to coherent states of photons.  This means that we do not expect a well defined number of photons to exist in the medium.  However, we can consider the case of a single photon entering the medium and ask what is the expected result.  We know that a fraction of the photon must now be absorbed into energy of the oscillators.  This suggests that the situation is one of a superposition of a system in the zero photon and single photon sector.  The waves between the plates are represented by right and left moving plane waves obeying $\omega=ck$ and, although the get the phase shifts at each plane we need a broad distribution of frequencies, we can use that the energy density to momentum density in each right and left moving component is constant by the quantization conditions $E=\hbar\omega$ and $p=\hbar k$.  The occupancy of the single photon sector is therefore given by $\mathcal{E}_{\text{em}}/\mathcal{E}_{\text{net}}$.  The remaining momentum and energy is in some excited state of the medium in the zero photon sector being distributed among various phonon and electronic excitations.  The extremely formal nature of quantum field theory and quantum optics generally preclude the kind of local detailed balancing analyses using conservation laws that classical systems allow.  This seems to be a nice exception and hopefully a step towards a richer set of cross checks on problems with quantum response in media.

\subsection{Stress Tensor}

The utility of the stress-energy tensor is in generating forces or, in the case of general relativity, providing sources for gravity.  The bulk stress tensor in the absence of a medium is often of questionable value since when we consider gradients of stress to get forces there must be something to push against.  This is the problem with idealized ``photon-gas'' hydrodynamic models in cosmology \cite{MTW}.  The local photons just travel ballistically regardless of whether or not they are in thermal equilibrium.  If there is a medium present then there must be sufficient time to equilibrate and transfer those forces to the material part of the medium and for the medium to reestablish equilibrium with it (either equilibrium with the charges as in our continuum model or thermal equilibrium).  The photon-photon interaction is so weak that we probe material changes, typically with other photons, to determine the response.  

The stress-energy tensor can be decomposed as 
\begin{align}
T^{\mu\nu} =T^{\mu\nu}_{\text{em-rad}}+T^{\mu\nu}_{\text{em-restoring}}+T^{\mu\nu}_{e}+T^{\mu\nu}_{\text{cores}}
\end{align}
where, for the progressive wave in the above coordinates,
\[ T^{\mu\nu}_{\text{em-rad}} =\left( \begin{array}{cccc}
\mathcal{E}_{\text{rad}} & \mathpzc{p}_{\rightarrow}+\mathpzc{p}_{\leftarrow} & 0 & 0\\
\mathpzc{p}_{\rightarrow}+\mathpzc{p}_{\leftarrow} & 2c|\mathpzc{p}_{\leftarrow}|& 0  & 0\\
0 & 0 & 0 & 0\\
0 & 0 & 0 & 0\end{array} \right)\]  
The force responses we usually desire involve the absorption of electromagnetic momentum by the medium itself at the boundaries and during propagation in the bulk where field's averaged behavior is changing.  Since the rate of transfer depends on both the opacity of the medium\cite{opacity} and response rate of the charges, it is hard to see how a collective stress-energy tensor for the combined system is of much value for this purpose.  Furthermore, there can be other sources of field induced stress in more general media due to changes in the mean electronic structure and nonradiative fields.  In this sense our dielectric is ``minimal'' in that the only forces that arise come from the radiative field and conservation laws for it.   Nevertheless, in the local frame of the medium, each component is simple, well defined and local conservation laws hold explicitly.  

\subsection{Microscopic Details of Periodic Solutions}
\begin{figure}
   \centering
   \includegraphics[width=3in,trim=10mm 30mm 10mm 110mm,clip]{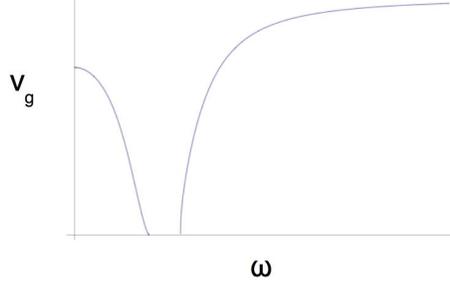} 
   \caption{Group velocity $v_{g}(\omega)$}
   \label{vg}
\end{figure}

The dispersion relation $\omega(k)$ gives us (fig.\ \ref{dispersionrelation}) how the frequency $\omega$ relates to the macroscopic wavevectors $k$.  Microscopically, however, the waves in the gaps are right and left moving combinations of wave with $k'=c/\omega$.   The group velocity is the mean velocity of motion of a packet and is plotted in fig.\ \ref{vg}.  We see from the dispersion relation eqn.\ \ref{dispersion} that there are two frequencies where the macroscopic $k$ vectors diverge, $\omega_{l}= \sqrt{\frac{K}{m}}$ and $\omega=\infty$, however now we see that only the first second one contains arbitrarily short microscopic wavelengths.  The second critical frequency, $\omega_{h}=\sqrt{\frac{K}{m}+\frac{\rho {e}^{2}}{m\epsilon_{0}}}$,  has $k=0$ but these are again made of microscopic waves of $k'=c/\omega_{h}$.  

\begin{figure}
   \centering
   \includegraphics[width=3in,trim=10mm 90mm 10mm 100mm,clip]{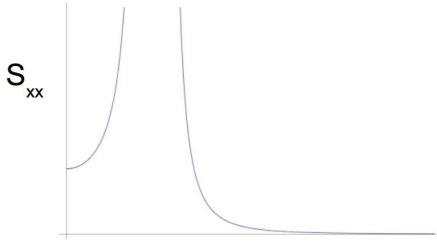} 
   \caption{Stress $S_{xx}(\omega)$}
   \label{stress}
\end{figure}

We can consider the internal stress of the medium in fig.\ \ref{stress} and we see that it diverges at both these critical frequencies.  This implies the internal standing waves are very large.  At $\omega_{l}$, the resonant frequency, we see that the internal energy diverges (for finite field strength) but $\mathcal{E}_{q}/\mathcal{E}_{\text{em}}$ also does so that we can say the system is ``elastically dominated.''  The other divergence seems to be an artifact of averaging fields since the microscopic $k$ goes to zero and both energies are finite.  If we compute it in terms of the internal field strengths the stress is actually finite.  In the case of metamaterials one can have a ``negative index of refraction'' so that the group and phase velocities move in opposite directions.  This analysis reassures us that, microscopically, the waves are all primarily moving in the same direction as the groups.  

%Many approaches to the microscopic analysis of the forces and stresses in a medium use a kind of dispersionless model.  This allows the group and phase velocities to be proportional.  Clearly they must have the wrong high frequency asymptotics.  We have seen in the above analysis that the subtleties of the surface impulses, internal stresses and equilibration impulses have required a distinction between these.  Furthermore, a full set of nonlinear branches are required to have the freedom to express the full initial data of the system.  The microscopic picture of electrodynamic waves near the resonances would not be apparent without the nonlinearity in the dispersion relation.  For this reason, a dispersionless analysis of waves in media is always inconsistent.  

\section{General Dielectrics}\label{damping}
\subsection{Damping and Causality in a General Medium}
So far we have only studied a nondissipative case of a simple medium with one resonance.  We could easily add other oscillators to get more complicated dispersion relations but they will always give singular behavior at the resonances.  Damping rounds these peaks and this is generally described by an imaginary part in the dielectric function.  Since we are seeking a description directly in terms of the microscopic motions instead of constitutive laws, we won't seek a direct analog of the dielectric function.  We are also only interested in using real quantities since we want an easy generalization to nonlinear and time changing media. 

If one specified a general real dielectric function one cannot even be sure that the imaginary part of the analytic extension will not include source terms (``negative damping'').  This is certainly problematic from the point of view of justifying causal evolution.  One could simply give a set of frequencies and damping rates for every function $\cos(k x)$ and enforce a linear evolution however, on Fourier transforming it, one generally finds that the damping function is not local.  This locality condition is what enforces the Kramers-Kronig relations for linear solutions.  Causality arises from these relations and the various precursors arise \cite{Brillouin} as higher velocity components that the rest of the packet during the ring-up time.  The usual treatments of causality involves some impressive complex analysis and one might wonder why this should be so.  Below we will discuss how conservation laws and positive definiteness of the energy density enforce this easily.  There is an analogous causality problem in the case of heat conduction.  Since this has led to a number of erroneous attempts at modifying the heat equation including higher order modified or extended schemes that give hyperbolic solutions (in violation of the fundamental degrees of freedom of the system) we include a discussion in the appendix based on a similar microscopic analysis in app.\ \ref{DiscreteW}.

A detailed description of the waves in the gaps showed there is a discontinuity in $B$ due to the currents at the plates but not in $E$.  (The electric field of the right propagating part of the field is illustrated in fig.\ \ref{wavesteps}. Including the hidden standing wave contribution between the gaps joins the net fields smoothly.)  For a wave with a well defined $\omega$ 
the component waves in the gap region obey the relation $\omega=ck$ exactly.  The plates in this limit now just look like reflectors and energy storage devices.  We saw that the distinction between traveling (progressive) and standing waves was that the relative phase of the magnetic field is different relative to the electric one.  This suggested that a deviation in the magnetic field due to $J$ in the previous model is actually introducing a partial reflected wave in the plate gap region.  If we initiate a pulse at one plate it causally travels to the next, partially reflects and transmits just as we expect from scattering theory.  

The continuum approach has too many advantages in its economy and analytic tools to abandon yet we must choose one that is not averaging away crucial information and retains some intuitive properties connecting it to the microscopic physics.  In analogy with the heat equation, it would be nice if we could place some limits on the velocity of the energy flux through the form of the equations themselves.  This would 1.\ manifestly  conserve energy and 2.\ exploit that energy is a positive density function so its vanishing ensures all the other variables vanish at a packet edge.  

The electromagnetic part of the dynamics are fixed by Maxwell's equations.  Since we have developed a model where $E_{\text{rad}}$ and $E_{\text{restoring}}$ are nicely separable that worked so well in elucidating the role of conservation laws, we assume that such a decomposition is generally valid or that there is some kind of universality that allows such a model to be equivalently constructed to every realistic medium.  The field evolution equations of $E=E_{\text{rad}}$ and $B=B_{\text{rad}}$ are then governed by the microscopic Maxwell's equations.  The interesting parts of the equations that involve ring-up, ring-down of the medium, damping, nonlinearity, time changing media response, etc.\ are then all in the details of the medium that manifest as the current and displacement functions.  
\begin{align}
\partial_{t}E&=c^{2}\nabla\times B-\epsilon_{0}^{-1}J_{e}\\
\partial_{t}B&=-\nabla\times E-\epsilon_{0}^{-1}J_{m}
\end{align}
where we have included a magnetic current $J_{m}$ as a device to create magnetic polarization later without having to explicitly use dipoles or current loops.  

Since $\nabla\cdot E_{\text{rad}}=0$ we have the wave equation for $E$ (and similarly for $B$)
\begin{align}\label{law}
\square E=\partial_{t}^{2}E-c^{2}\nabla^{2}E=\frac{1}{\epsilon_{0}}\left(-\nabla\times J_{m}+\partial_{t}J_{e}\right)
\end{align}
The fields then propagate at the speed of light with source and sink terms as long as $J_{m}$ is not a function of $\nabla E$ and $J_{e}$ is not a function of $\partial_{t}E$.  This then preserves the characteristic structure of the equations.  

In a realistic medium there is an elastic constant for the electron distortions, $K_{e}$, and one for the cores, $K_{c}$.  The electron distortion is measured relative to the cores so that these are coupled.  Rapid oscillations will tend to drive the electrons and leave the cores behind as in the case of internal black/gray body radiation.  Bulk elastic forces, as $P_{S}$, that change slowly will transfer directly to the cores and create acoustic effects.  However, the ability of the medium to store and deliver energy through the currents can be quite general and still maintain causality.  We can let $J_{e}$ and $J_{m}$ be local functions of the fields but also of elastic and other local properties of the material and even of external driving forces.  The simple linear case of a dielectric gives $\dot{J}_{e}=\frac{q^{2}}{m}E-\frac{K q}{m}Y$ where $J_{e}$ itself is an independent variable so no constitutive law holds for it in terms of the fields.  Rapid damping of acoustic modes will tend to lead to relaxation to the optical modes in eqn.\ \ref{dispersionrelation} so that an apparent constitutive law is observed.  

As long as the local medium dynamics are a function of the local medium effects and not derivatives of the fields, causality is manifestly preserved.  The medium will tend to damp deviations from the optical modes through faster damping losses in the acoustic modes.  These lead to the precursors that absorb electromagnetic energy as a packet evolves.  The natural time scales associated with the medium that determine how fast energy is absorbed from a nearly monochromatic wave that is not yet in equilibrium with the medium are given by the rate of ring-up to an equilibrium value of $\mathcal{E}_{q}/\mathcal{E}_{\text{\text{net}}}$.  

As an example, consider the case of the free electromagnetic wave as a monochromatic beam in a dielectric medium with no initial medium response.  The source terms must conspire to give advance at the phase velocity so that
\begin{align}
\partial_{t}^{2}E(\omega)-c^{2}\nabla^{2}E(\omega)&=\epsilon_{0}^{-1}\partial_{t}J_{e}\\
&=(-c^{2}+v_{\text{ph}}^{2}(\omega))\nabla^{2}E(\omega)
\end{align}
Deviations from this situation as in the case of a free space EM wave inserted in the medium with no medium response yet present, allows the fields to advance at $c$ while the medium gradually steals energy and momentum from the beam.  Such a configuration can be expressed on the basis set using the four branches of the dispersion relation in eqns.\ \ref{B}-\ref{solution1}.  For an infinite wavetrain, the wavelengths do not change but the frequency is altered from the free space dispersion relation, the amplitude decreases, and backwards components are generated.  The resulting four-wave mixing, $\pm\omega_{a}(k),\pm\omega_{o}(k)$ gives an oscillatory change in the fraction of energy storage in the electric field.  

For an advancing packet, there is a broad spectral distribution of Fourier components.  The Sommerfeld theory \cite{Brillouin} of precursors states that each component advances at approximately the group velocity of it.  In real space, the local absorption of EM energy from a gradually sloping monochromatic packet by the medium can be described with current such that $\partial_{t}J$ gives the RHS of
\begin{align}
\partial_{t}^{2}E(\omega)-c^{2}\nabla^{2}E(\omega)&=(-c^{2}+v_{\text{ph}}^{2}(\omega))\nabla^{2}E(\omega)+U
\end{align}
where $U$ is approximately antiparallel to $\partial_{t}E$.  

The most general medium is described by a set of variable $M_{1},M_{2}\ldots M_{N}$ with equations of motion $\dot{M}_{i}=h_{i}(\{M_{j}\},E,B)$.  The displacement and currents $Y, J$ are functions of the $M_{j}$.  Electromagnetic restoring forces are hidden in the material variables so that electrostrictive effects can arise in these equations.  Additionally, external forces and sources of energy can be injected that can change the properties of the material, as in the case of electromagnetically induced transparency, or chemically driven medium changes.  Therefore $Y,J$ can be explicit functions of time as well.  The momentum effects on the medium can either be computed exactly through the local Lorentz forces or by conservation laws utilizing that the form of the Poynting vector in the medium is unchanged from that of the vacuum.  We can now view eqn.\ \ref{law} as a local real-space and manifestly causal set of equations of motion where the local evolution of $J_{e}$ and $J_{m}$ are functions of local medium conditions and the local values of $E$ and $B$.  

We can think of this problem as a balance of energy in the equilibrium case where the input energy from the fields $E\cdot J$ is balanced by the output $\nabla\cdot S$ for each charge layer.  Just as in the radiation reaction case of a point particle \cite{Rohrlich} there are necessary nonlinearities here to give the right damping modification of the the driving force that arise from the relativistic acceleration.  For larger field strengths these are unavoidable as the medium response gains nonlinear changes during ring-up and ring-down even if the elastic restoring forces stay in the linear regime.  

The extensions of linear response theory are typically rather formulaic.  One imposes a structure based on a hydrodynamic framework and seeks relativistic or nonlinear corrections consistent with some physically important symmetries.  The relativistic correction is not extremely important since it takes enormous fields to drive electrons at relativistic speeds and producing relativistic changes in medium velocity over field equilibration distances is similarly difficult.  The whole constitutive approach has locality implicitly involved since it assumes the fields and medium reach an equilibrium over distances short compared to those of physical interest.  A correction to medium response theory that seems more relevant is in how deformation and noninertial effects on the medium that may be small but potentially iterative many times over an optical path may build to produce large net effects.  A medium undergoing acceleration will have the bound radiators move and reradiate without a time lag compared to the radiation that is momentarily unbound from it.  Such a treatment seems essentially nonlocal and to require the kind of explicit decomposition of fields and medium we have discussed here.  

\subsection{Universality}
There have been many attempts at determining the kinds of forces and stresses in dielectric media with some of the opinion there is a unique decomposition in terms of the electromagnetic and media response and others arguing that such a decomposition is meaningless and correct use of boundary conditions give equivalent results.  The model introduced here, clearly gives a unique decomposition and leads to the hope that by tracking the flux of energy and momentum of a traversing radiation field and considering the absorption ring-up and damping of the medium one could arrive at a universal picture of dielectrics with a unique decompositions of field and medium energy and momentum.  In the case of densely packed atoms in condensed matter, the delocalized electron wavefunctions give a momentum that is a linear combination of $A$ and $\nabla\phi$, the phase gradient of the wavefunction, this is explicitly gauge dependent so this is problematic already.  

\begin{figure}[h]
\centering
   \includegraphics[width=5in,bb=0 0 1000 1000]{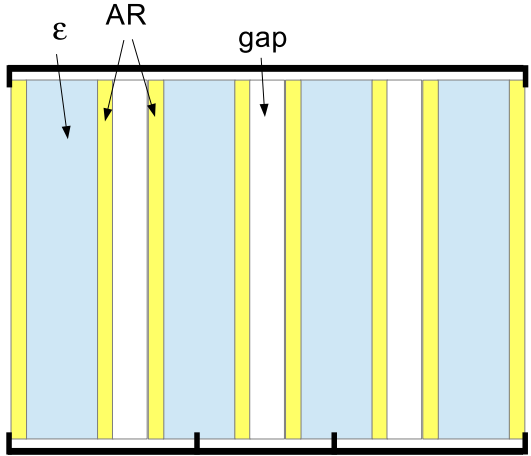} 
   \caption{Antireflective cladded dielectric slabs held by point contacts on a frame.}
   \label{nonuniversal}
\end{figure}

We can consider a medium made up of layers of independent dielectric blocks with AR coatings so that no standing wave exists between the layers as in fig.\ \ref{nonuniversal}.  These subblock layers will have internal stress but the walls of the net medium have none.  This then gives a kind of metamaterial with nontrivial dielectric response but no stress which seems to kill our idea of such general universality.  A second problem is the long range polarization forces that exist at the edge of beams or blocks of media.  These fields can fall off rather slowly and can create local stretching and long range attraction that then couples to the elastic response of the medium.  For long wavelengths these forces seem to involve the nonradiative fields of the medium.  The injected radiative field is the only source of energy in the problem so must lose energy to fund such additional energy expenditures.  The dispersion relation was shown to be a measure of how much energy is stored in the medium versus the field so must be impacted by such additional processes.  Even in the linear limit, one can have long range effects that are easy to neglect in a ``constitutive model.''  An example is the long range electrostatic forces on DC wires from the gradient of surface polarization to drive the internal small but nonzero electric field.  

It is unclear how much such considerations impact the dielectric response of more mundane materials that have not undergone some clever small scale engineering.  However, it does suggest that universality does not exist generally for the stresses in dielectrics as function of the medium's dielectric and permeability functions.  Furthermore, even in the apparently linear regime for medium response one may need to do more careful detailed balancing of the internal fields and elastic medium response than can be obtained by use of a naive stress tensor as a function of internal field strengths.  

\section{Conclusions}
An alternate title for this article could have been ``Electrodynamics in Media without Constitutive Laws'' in that we provide a full dispersion relation so that fields and charge motion may be chosen with complete independent freedom.  We have presented a simple dielectric model that is easily extendable to the case of multiple resonances and that does not depend on any averaging over localized oscillators and fields.  The forces and impulses are now readily apparent by microscopic analysis in terms of the purely radiative field strengths that are not always clearly separable from the restoring fields in the averaging approaches.  Some nice byproducts of this have been a complete way to track momentum in the entire system, a measure of the photon number changes as a quanta enters and traverses the media, some understanding of how acoustic losses can be generated and an easy way to see how group velocity relates to the hidden microscopic phase velocity in metamaterials.  

Nonlinearity and time delay are two possible routes to chaos and at higher intensities and frequencies, media response has both of these.  As our ability to generate higher field intensities grow, more exotic dynamics are almost certain to arise.  Additionally, as our ability to measure smaller effects improve, the necessity to describe optical impulses and other competing small effects becomes more important.   In addition to the case of large bond distortion, the Abraham-Lorentz-Dirac force law ensures that nonlinearity will arise as a function of field gradients.  The finite granularity scale of media gives a retardation for propagating changes to advance between separated radiators and to relax to the new mean frequency and field energy.  In the case of nonlinear media one can seek a modification of the Kramers-Kronig relations.  Generally, a local damping law is assumed but phonons mediate most losses and these are nonlocal objects and we are nowadays frequently concerned with confined systems where discreteness in the phonon spectra is important.  It therefore is advantageous to be able to discuss damping and dielectric response more generally than the linear response method and more intrinsically than a formal nonlinear extension of it.  

Granular materials provide the canonical example of a system with no useful continuum limit.  Hydrodynamics does not work for granular flows.  Static packings are strongly history dependent and focus forces over many orders of magnitude at the scale of individual grains.  Frictional torques produce an ``indeterminacy'' whereby the packing and its boundary forces are inadequate to determine the internal stress.  In the case of an electrodynamic field moving in media, one has to make the assumption that the fields are changing slowly on the scale of the separation of charges.  For chirped and radically shaped pulses this is not so.  X-ray and gamma ray shallow angle reflection will not satisfy this condition either.  Having an explicit model that can describe the fields on the scale where the wavelength becomes comparable or smaller than the granularity scale of the medium may give new ways to address such problems where continuum mechanical ideas may no longer be valid.  

Precursors have been difficult to detect at a level of accuracy to test the Sommerfeld-Brillouin \cite{Brillouin} theory.  For this reason people have been hunting for other systems e.g.\ surface gravity waves, for comparison \cite{Falcon}.  Metamaterials are highly tunable.  Resistance and real time variations can be easily introduced into the shaped oscillators for microwave frequencies due to their larger scale.  This would provide an excellent place to test for precursor shape and damping effects and introduce controls on the flux they generate.  Just as importantly, a clear understanding of the microscopic reality may help weed out some of the more fantastical theories and flawed analogs to validate them.

There are other examples of systems with natural velocity limits that are not $c$.  The speed of ocean waves, the motion of oscillations on a spring or heat transport in a solid have natural limits in the speed of sound of the underlying medium.  It is unclear if there is any universality to the resolution of the dynamics here.  Certainly some will seek a sweeping class of equations based on symmetry that ignore the underlying details of the dynamics.  The results here and similar ones not included here are suggestive that this is a mistake.  Ultimately everything comes from microscopic physics and shortcuts that seem initially successful or are only successful in particular cases can contribute to long lasting confusion.  

The quantum use of quasiparticles is widespread yet the form of their dispersion relations imply that they must face such similar constraints.  The best form for such corrections is an important challenge.  High temperature superconductivity comes from a strongly interacting system utilizing very shallow hole filling bands.  It seems that one can have extra energy carried with electrons in the interactions to give larger effective mass but how $m^{*}<m_{e}$ can arise seems mysterious.  This is only possible due to the fact that the electron-electron interaction is greater at the Brillouin zones so that excitations can actually reduce the net interaction.  It would be interesting to see if such a composite model of electron waves modeled on the example given for electromagnetic waves here might lead to new insight on the forces and transport in conducting media.

\appendix

\section{Left Handed Materials}
A current exciting topic in optics concerns the so called ``left handed'' or ``negative index'' materials \cite{Veselago, Pendry}.  In this case one has a group velocity that is opposite to the phase velocity.  The electric, magnetic and propagation vectors then become a left handed pair.  Such a state is clearly not possible for a simple dielectric.  It it therefore necessary that the medium's dielectric and permeabilities are both nontrivial.  In the case that they are both negative such a condition exists.  It is therefore interesting to see how such a state can be made sense of microscopically in terms of the model of free radiation fields traveling between oscillators that act as temporary energy and momentum storage devices.  Such materials have been constructed in the microwave regime with ``split ring'' lattices to get both electric and magnetic responses that are coupled.  Such devices can be modeled with the more general case of electric and magnetic monopoles as oscillators as was used in Sec.\ \ref{AR}.  Although such magnetic monopoles may not exist in nature they can model any current loop's action and do so in a parallel and simple fashion of elastically bound massive monopoles.  

\begin{figure}[h]
\centering
   \includegraphics[width=5in,bb=0 0 1000 1000]{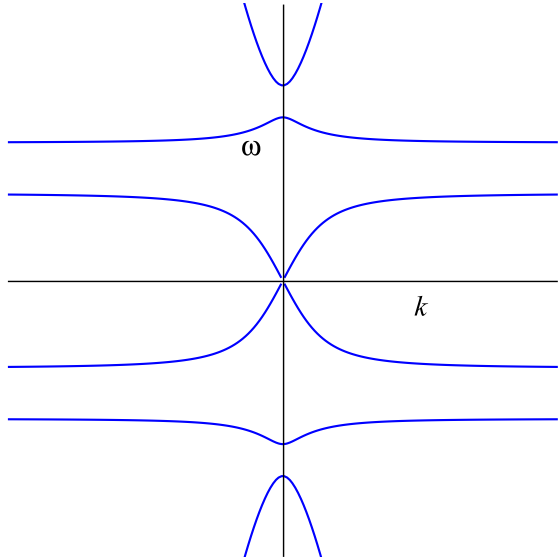} 
   \caption{Dispersion relation for electric and monopole oscillators. }
   \label{lefthanded}
\end{figure}

The resulting equations of motion are
\begin{align}
\nabla\times B&=\mu_{0}(\rho_{e}\partial_{t}Y_{e}+\epsilon_{0}\partial_{t}E)\\
-\nabla\times E&=\epsilon_{0}^{-1}(\rho_{m}\partial_{t}Y_{m}+\epsilon_{0}B)\\
\it{m}_{e}\partial_{t}^{2}Y_{e}&=\rho_{e} E-\kappa_{e}Y_{e}\\
\it{m}_{m}\partial_{t}^{2}Y_{m}&=\rho_{m} B-\kappa_{m}Y_{m}
\end{align}
where $m_{e}$ is the charge density of the electric oscillators, $m_{m}$ is the monopole density of the magnetic oscillators, etc.  
The dispersion relation is 
\begin{align}
k^{2}=\frac{\omega^{2}}{c^{2}}\left(1+ \frac{\rho_{e}^{2}}{\epsilon_{0}(\kappa_{e}-\omega^{2}m_{e})}\right)\left(1+ \frac{\rho_{m}^{2}}{\epsilon_{0}(\kappa_{m}-\omega^{2}m_{m})}\right)
\end{align}
as an obvious extension of the results for the purely dielectric case in eqn.\ \ref{dispersion}.  
A typical solution is shown in fig.\ \ref{lefthanded}.  Comparing with the purely dielectric dispersion relation, we see that presence of two additional bands.  The second from the bottom gives negative $v_{ph}$ and positive $v_{g}$ consistent with the properties of left handed materials.

Interestingly, the energy flux between the plates must be advancing as free photons in the group velocity direction.  The advancing phase direction is for the spatially averaged wave.  In our model this can be represented as free waves with phase shifts at each oscillator layer as in fig.\ \ref{wavesteps}.  For a dielectric with no magnetic response, this can be viewed as the magnetic field intensity or the component of the electric field intensity that is purely propagating.  For our negative index material, it should be viewed as the purely propagating component of either field.

\section{Discrete Walkers}
\label{DiscreteW}
Consider a 1D lattice of points separated by $d$ that each contain $n(i)$ walkers.  Every discrete time increment $\tau$ the system is updated and half of the walkers move left and half right one step.  (We assume $n$ is always so large that problems posed by odd values do not make important contributions.)  There is no a priori reason $n(i)$ should be quasicontinuous but, assuming it is, we can modify the finite difference equation into an approximate continuum one:
\begin{align}
n(i,k+1)&=\frac{1}{2}(n(i-1,k)+n(i+1,k)) \Rightarrow\\
n(x,t+\Delta t)&=\frac{1}{2}(n(x-\Delta x,t)+n(x+\Delta x,t))\Rightarrow\\
\dot{n}(x,t)&\approx\frac{1}{2}\left(\frac{\Delta x^{2}}{\Delta t}\right)n''(x,t)=Cn''(x,t)=\frac{d}{dx}j(x,t)
\end{align}
where $j=C\frac{d}{dx}n$.  In the last line we reformulated this into a form reminiscent of $\dot{q}=\nabla\cdot j$ where $j$ is the current of the scalar quantity $q$.  

First note that the discreteness of the system places bounds on the possible gradients of $n$.  Since $n$ is a positive definite quantity $j/C=\frac{d}{dx}n\le n/\Delta x$.  Initial data should remain less than this and the evolved function should preserve this condition.  
From here we could attempt to improve the accuracy of the equation by using higher order terms in $t$ and $x$ derivatives.  Higher order terms in $t$ violate the sufficiency of $n(x)$ as initial data.  We could attempt to remove these terms by some iteration of lower order approximations to get self consistency (as it done with the radiation reaction (see LL)).  It is doubtful if any finite number of spatial derivatives would accomplish the goal of keeping the evolution of the edge bounded by the velocity $u=\Delta x/\Delta t=2C/\Delta x$, as is true for the finite difference equation, and any localized initial data.  Infinite order equations are unwieldy and can be argued to be disguised nonlocal equations.  

To look at this from a different perspective, consider the current $j$ and what it tells us about the velocity of the propagation.  We can decompose this current in to right and left moving components $j(x)=j_{+}(x)+j_{-}(x)\approx (-C/\Delta x)n(x-\frac{1}{2}\Delta x)-(-C/\Delta x)n(x+\frac{1}{2}\Delta x)$.  We know from above that the velocity of the flux in each direction is $v_{\pm}=\pm\frac{\Delta x}{\Delta t}=\pm2C/\Delta x=j_{\pm}/n_{\pm}$ where $n_{\pm}=\frac{1}{2}n$ is just the fraction of n from a site that moves right or left respectively.  By the above physicality condition we have $|j/n|\le C/\Delta x$.  If the solution approaches this we know we have moved into unphysical territory i.e.\ $n(i)$ is ``very large'' but an adjacent value is now approaching zero.  This means the flux from one side is effectively terminated and increasing $n''$ does not increase the flux beyond $n_{\pm}u$.  We can best modify Fick's law by introducing a scale dependent diffusion constant that drives the diffusion to zero as a function of $(j/n)^{2}-(C/\Delta x)^{2}$.  
\begin{align}
\dot{n}(x,t)&=\frac{d}{dx}(\tilde{C}n'(x,t))
\end{align}
where $\tilde{C}=Cf((C/\Delta x)^{2}-(j/n)^{2})$.  $f(s)$ can be as simple as a step function $\Theta(s)$ or a smoothed version that still vanishes at $s=0$.  This introduces some essential nonanalyticity  and nonlinearity in the problem.  This is the price we have to pay to preserve the possibility of having a continuous and smooth set of descriptors to evolve the system with the correct dynamical degrees of freedom.  In the case $f=\theta$ we can immediately see that causality is preserved because the evolution is just the heat equation until the local velocity reaches $u$ when it halts until the nearby function changes enough so that it can again evolve at a lower speed.  In practice this state is never reached if $\theta$ is rounded over to a smooth function.  
As a final note, we see energy is conserved both globally and locally by Gauss's law.  

It still remains to find a microscopic derivation of the function $f(s)$ or even to see what the form of corrections to this model might be.  This will have to remain to future work.  To experimentally measure this quantity, using the edge of a packet of heat seems extraordinarily difficult.  One could probe rapidly brought together surfaces with large temperature differences.  For diffusion, one could look for rapidly moving outliers and large number averages.  

Nonlinear equations follow as descriptions of truly linear phenomena when we suppress dimensionality e.g.\ the Schr\"{o}dinger equation and Hartree-Fock, density functional theory and the Gross-Pitaevskii equations.  In the case of quantum field theory, nonlinearities show up in the running of the coupling constant as a result of suppressing scales that correspond to energies beyond what is physically relevant for the given problem.  
Here we have shown that nonlinearities follows from neglected (and apparently intractable to the analytic tools of smoothing and taking limits) small scale and discrete physics when we need to work with a best fit continuum model.  

It is interesting to note that equations like the porous medium equation $\dot{n}=\nabla\cdot(n^{m}\nabla n)$ with $m\ge1$ give a set of weak solutions (essentially continuous solutions with smoothness discontinuities) that have finite velocity of front propagation.  We can see that these also have a vanishing current where $n$ approaches zero.  In contrast we have sought a solution where the velocity of propagation is bounded by a physically fundamental quantity.  If we were to extend this to heat transport in a solid we note that the phonons are bounded by the velocity of sound.  As long as we are at temperatures where the relevant phonons are from the linear part of the dispersion relation, we can expect such an equation to be relevant.   The gas of heat flow in gases is more complicated.  The particles can have a very broad distribution of velocities.  This above result would have to use contribution over bounds from all these velocities and the sharp edge we see above would become blurred.  Even though sound speed is closely related to the thermal mean velocity the ``thermal edge'' of the heat distribution would creep out beyond it.  The speed of light bound will be established once a relativistic distribution of particle speeds is used.  

\section{The Stress Tensor and Momentum}\label{StressMomentum}

The stress-energy tensor of a system $T^{\mu\nu}$ is often described as the flux of $p^{\mu}$ momentum across the $\nu$th hypersurface composed of normals to the vector $x^{\nu}$.  This is a common approach in many texts on continuum mechanics.  Since we are interested in the microscopic interplay between material and radiation in a way that treats neither as the ``driver'' or ``responder,'' as is generally done in derivations of the Kramers-Kronig relations, it is good to pause and look at some specific cases from a microscopic point of view.  

In the case of a gas, the above is a good model.  All the pressures and stresses in the medium are the results of kinetics.  As such, these are the direct result of microscopic transfer of momentum.  The case of solids and liquids are different.  There is a kinetic component but there is also potential energy in the bonds between atoms that can be strained and do the work of transferring forces across the parcels.  This may seem academic but when we look at some formulations of momentum conservation in continua, we can start to consider the naively implied ``momentum flux'' from such a $T^{\mu\nu}$ as though it is a conserved quantity.  Specifically $\partial_{i}T^{ij}=f_{i}=-\partial_{j}P$ implies a \text{globally} conserved momentum (since $\int\partial_{j} P dV=0$) however the \text{local} conservation law $\partial_{t}(\rho v^{j})+v^{i}\partial_{i}(\rho v^{j})=0$ (or $\dot{\vec{p}}+\vec{v}\cdot\nabla\vec{p}=0$) is not generally true even when external forces are zero. (Note that for constant density fluids $\partial_{i}v^{i}=0$ we can rewrite this as $\partial_{t}(\rho v^{j})+\partial_{i}(\rho v^{i}v^{j})=0=\partial_{i}\Pi^{ij}$ for $\Pi^{ij}=\rho v^{i}v^{j}$.  This is the correct Eulerian momentum flux in the lab frame assuming any microscopic motions can be neglected.  {We generally refer to these fluids as ``incompressible'' but one can have fluids with varying density e.g.\ from varying solute concentration where each parcel is effectively incompressible so ``constant density'' is more accurate here.})  
 This is what we would expect for a conserved momentum \text{flux}.  Instead we see that $\nabla P$ plays the role of a source and sink term for it.  Momentum is exactly conserved and, absent any electromagnetic contribution, is the same as the mass density flux.  The subtleties with momentum flux is that it is not locally conserved and that one often tries to write forces as momentum fluxes that may  partially cancel on a given material parcel.  During a microscopic evaluation of the sample one only sees bond strain and net motion.  The net result is the same net force but momentum flux in this sense has no microscopic meaning.  

If we consider a gas we can locally keep track of the momentum of the system and how it shifts at each collision.  In fact we can give a locally well defined right and left moving momentum flux.  In the case of condensed matter, this is not the case and the elastic forces driven by the potential energy must be included in the microscopic description of the stress.  Inspecting a stressed zero temperature solid microscopically will indicate the strained bonds but no momentum is being transferred.  

When the fields exist in the presence of media we can lump the net local stress into a stress tensor.  It is not clear how valuable this is since the waves might propagate through the medium for some distance before it relaxes.  We generally try to use such stress tensors to take gradients to derive local forces.  These must be the net forces on particles and fields.  
If this induces reflection and packets do not evolve as we expect from linear combination of the states in a dispersion law it is not clear how much each one responds to this.  

When electromagnetic fields are being included, we need to include its momentum.  How to decouple these from media is not entirely obvious for realistic media.  Several respectable sources have promoted some confusion about the nature of the ``canonical'' momentum of a particle in an electromagnetic wave \cite{LL} and this ends up playing a role in some discussions of the meaning of the Abraham and Minkowskii momentum in media \cite{Barnett}.  
Since it is easy to clarify we shall do it here.  

The canonical momentum of a particle is stated as $p^{i}=mv^{i}+\frac{q}{c}A^{i}$. (Gaussian units are used in this paragraph.)  There are attempts to use this and other pseudomenta to give true forces at gradients or boundaries.  Some discussion has taken place as to when this is valid \cite{McIntyre}.  Let us first consider the gauge problem and this proposed definition.  Clearly it is not invariant.  We know that these problems go away in the case of the Schr\"{o}dinger equation because the phase of the wave transforms to cancel any vector potential gauge change.  In this case we have
$m j_{i}=p_{i}=\rho({\hbar}\partial_{i}\varphi+\frac{q}{c}A_{i})$.  The gauge transformation of $A_{i}$ is now compensated by a stretching in the phase, $\varphi$, of the wavefunction.  By Ehrenfest's Theorem, a localized packet moving with with velocity $v$ continues along the Lorentz force law path.  
To get the original canonical relation to give the true momentum we need to choose and maintain a gauge choice of $A_{i}=0$ at each particle.  
The wavefunction packet constitutes a model for the classical charged particle to the extent the (in this case, irremovable) self-forces can be neglected.  Since $mv$ is a gauge invariant quantity, the classical canonical momentum really only coincides with this gauge restricted case.  In this model, $v$ is a function of the phase gradient of the packet and the particular value of $A^{i}$ chosen for the given  magnetic field.  
Any modification of this vector potential by a gauge choice $A^{i}\rightarrow A^{i}+\partial^{i}\Phi$ should leave $j^{i}$ unchanged and give no new forces on the particle.  This implies that the only classical ``canonical'' transformation of the momentum that should be allowed obey the same local constraints i.e.\ $\partial^{i}\Phi=0$ at each charge.  

\section*{Acknowledgements} The author acknowledges helpful feedback from David Aspnes.


\begin{thebibliography}{}
\bibitem{note} We properly should consider the elastic coupling of the electrons to the cores and the cores to each other as distinct.  The time scale of these oscillating forces and responses determine how much momentum is taken up by electrons alone versus electrons and cores together.  This creates a distinction between a plasmon-like and acoustic modes.  These vibrations will create an elastic stress on the bonds, hence the medium, despite having no net longitudinal momentum contributions.  The inclusion of such modes will increase the number of branches in the dispersion relation.
\bibitem{opacity} The model we have here of uniform plates will be replaced by separated localized oscillators in realistic media.  These are characterized not only by mass, charge and elastic response but cross section.  This allows for longer distances to equilibrate changes in field strength and frequency of the propagating waves for a fixed average $\rho_{e}$, $\rho_{m}$ and $\kappa$.
\end{thebibliography}
\end{document}